\documentclass[submission, Phys]{SciPost}

\usepackage{tikz}
\usepackage{braket}
\usepackage{xcolor}
\usepackage{float}
\usepackage{graphicx}
\usepackage{dsfont}
\usepackage[export]{adjustbox}
\usepackage{blkarray}

\definecolor{FAUblau}{RGB}{4,49,106}
\definecolor{FAUdunkelblau}{RGB}{4,30,66}
\definecolor{FAUgelb}{RGB}{253,183,53}
\definecolor{FAUorange}{RGB}{232,119,34}
\definecolor{FAUrot}{RGB}{197,15,60}
\definecolor{FAUdunkelrot}{RGB}{151,27,47}
\definecolor{FAUmedblau}{RGB}{24,180,241}
\definecolor{FAUmeddunkelblau}{RGB}{0,82,135}
\definecolor{FAUgruen}{RGB}{123,183,37}
\definecolor{FAUdunkelgruen}{RGB}{38,97,65}
\definecolor{FAUmetallic}{RGB}{140,159,177}
\definecolor{FAUdunkelmetallic}{RGB}{47,88,110}
\DeclareRobustCommand{\tcplaq}
{
	{\hbox{\begin{tikzpicture}[x=1 ex, y = 1 ex, scale = 0.25, style={inner sep=0,outer sep=0}, line cap = round, line join= round, line width = 0.8]
			\draw  (0,0) rectangle(3 ,3);
			\end{tikzpicture}}
	}
}

\DeclareRobustCommand{\tcstar}
{
	{
		\hbox{\begin{tikzpicture}[x=1 ex, y = 1 ex, scale = 0.25, style={inner sep=0,outer sep=0},line cap = round, line join= round, line width = 0.8]
			\draw  (0,1.5) -- (3,1.5) ;
			\draw  (1.5,0) -- (1.5,3) ;
			\end{tikzpicture}}
	}
}
\DeclareRobustCommand{\bigtcplaq}
{
	{\hbox{\begin{tikzpicture}[x=1 ex, y = 1 ex, scale = .35, style={inner sep=0,outer sep=0}, line cap = round, line join= round, line width = 1]
			\draw  (0,0) rectangle(3 ,3);
			\end{tikzpicture}}
	}
}

\DeclareRobustCommand{\bigtcstar}
{
	{
		\hbox{\begin{tikzpicture}[x=1 ex, y = 1 ex, scale = .35, style={inner sep=0,outer sep=0},line cap = round, line join= round, line width = 1]
			\draw  (0,1.5) -- (3,1.5) ;
			\draw  (1.5,0) -- (1.5,3) ;
			\end{tikzpicture}}
	}
}

\DeclareRobustCommand{\tcplaq}
{
	{\hbox{\begin{tikzpicture}[x=1 ex, y = 1 ex, scale = 0.25, style={inner sep=0,outer sep=0}, line cap = round, line join= round, line width = 1]
			\draw  (0,0) rectangle(3 ,3);
			\end{tikzpicture}}
	}
}

\DeclareRobustCommand{\tcstar}
{
	{
		\hbox{\begin{tikzpicture}[x=1 ex, y = 1 ex, scale = 0.25, style={inner sep=0,outer sep=0},line cap = round, line join= round, line width = 1]
			\draw  (0,1.5) -- (3,1.5) ;
			\draw  (1.5,0) -- (1.5,3) ;
			\end{tikzpicture}}
	}
}
\DeclareRobustCommand{\bigtcplaq}
{
	{\hbox{\begin{tikzpicture}[x=1 ex, y = 1 ex, scale = .35, style={inner sep=0,outer sep=0}, line cap = round, line join= round, line width = 1]
			\draw  (0,0) rectangle(3 ,3);
			\end{tikzpicture}}
	}
}

\DeclareRobustCommand{\bigtcstar}
{
	{
		\hbox{\begin{tikzpicture}[x=1 ex, y = 1 ex, scale = .35, style={inner sep=0,outer sep=0},line cap = round, line join= round, line width = 1]
			\draw  (0,1.5) -- (3,1.5) ;
			\draw  (1.5,0) -- (1.5,3) ;
			\end{tikzpicture}}
	}
}

\hypersetup{
    colorlinks,
    linkcolor={red!50!black},
    citecolor={blue!50!black},
    urlcolor={blue!80!black}
}

\usepackage[bitstream-charter]{mathdesign}
\DeclareSymbolFont{usualmathcal}{OMS}{cmsy}{m}{n}
\DeclareSymbolFontAlphabet{\mathcal}{usualmathcal}
\begin{document}

\begin{center}{\Large \textbf{
 Incorporating non-local anyonic statistics into a graph decomposition
}}\end{center}

\begin{center}
 Matthias Mühlhauser, Viktor Kott, Kai Phillip Schmidt\textsuperscript{$\star$}
\end{center}

\begin{center}
{Department of Physics, Staudtstra{\ss}e 7, Friedrich-Alexander-Universit\"at Erlangen-N\"urnberg, Germany}
\\
${}^\star$ {\small \sf kai.phillip.schmidt@fau.de}
\end{center}

\begin{center}
\today
\end{center}


\section*{Abstract}
{\bf
In this work we describe how to systematically implement a full graph decomposition to set up a linked-cluster expansion for the topological phase of Kitaev's toric code in a field. This demands to include the non-local effects mediated by the mutual anyonic statistics of elementary charge and flux excitations. Technically, we describe how to consistently integrate such non-local effects into a hypergraph decomposition for single excitations. The approach is demonstrated for the ground-state energy and the elementary excitation energies of charges and fluxes in the perturbed topological phase.
}

\vspace{10pt}
\noindent\rule{\textwidth}{1pt}
\tableofcontents\thispagestyle{fancy}
\noindent\rule{\textwidth}{1pt}
\vspace{10pt}

\section{Introduction}
\label{sec:intro}
The research on topological quantum phases \cite{Wen_1989,Wen_1990,Wen_2004} is a very dynamic field in modern physics. Such exotic phases have highly entangled ground states that cannot be described by Landau spontaneous symmetry breaking and possess anyonic elementary excitations with non-trivial particle statistics \cite{Leinaas_1977,Wilczek_1982} deviating from conventional bosons and fermions. These anyons are the essential ingredient for the field of topological quantum computing exploiting non-Abelian braiding properties \cite{Kitaev2003,Nayak_2008}. Topological order plays a crucial role in the physics of the fractional quantum Hall effect \cite{Laughlin_1983,Tsui_1982} and in frustrated quantum systems as quantum spin liquids in quantum simulators or condensed matter realizations \cite{Balents_2010,Savary_2017}. In this context, strongly correlated Mott insulators with strong spin-orbit interaction like the iridates \cite{Jackeli_2010,Singh_2012} as well as $\alpha$-RuCl$_3$ \cite{Plumb_2014,Banerjee_2017,Banerjee_2018} have been intensively studied in recent years. These quantum materials are discussed as potential realizations of Kitaev's honeycomb model \cite{Kitaev_2006}, which is exactly solvable possessing topologically ordered ground states. In the limit of strong anisotropic interactions Kitaev's honeycomb model can be mapped to the well-known toric code \cite{Kitaev2003} arising as an effective low-energy model in perturbation theory.

The 2D toric code represents an exactly solvable microscopic model. This model realizes a non-trivial topological phase with a highly entangled ground state and Abelian anyons as elementary excitations. As a consequence, it has been an attractive starting point for many investigations to understand the physical properties of topologically ordered quantum matter, e.g., the effect of external perturbations and the properties of induced quantum phase transitions \cite{Trebst2007,Hamma2008,Yu_2008,Vidal2009,Vidal2011,Tupitsyn2010,Wu2012,Dusuel2011,Iqbal2014,Morampudi2014,Zhang2017,Vanderstraeten2017}, the consequences of thermal fluctuations \cite{Alicki_2009,Castelnovo_2007,Nussinov_2009_b}, the properties of entanglement measures \cite{Halasz_2012,Santra_2014} or dynamical correlation functions \cite{Kamfor2014} as well as non-equilibrium properties \cite{Tsomokos_2009,Rahmani_2010}. Further, generalization of the perturbed toric code has been investigated in 3D \cite{Hamma2005,Nussinov2008,Reiss2019,Williamson2021,Delcamp2021}, in layered systems \cite{Wiedmann2020,Schamriss2022} and for frustrated geometries \cite{Schmidt2013,Tarabunga22}. 

One prominent method which has been applied successfully to determine the quantum-critical properties of the 2D toric code in the presence of a uniform magnetic field \cite{Vidal2009,Vidal2011,Dusuel2011} are high-order series expansions applying the method of perturbative continuous unitary transformations \cite{Knetter2000,Knetter2003}. In these works the excitation energies of elementary charges and fluxes have been calculated as a series in the low-field limit. The gap closing is then analyzed to extract quantum critical points and associated critical exponents. The non-local effects of the braiding statistics have been taken into account in a post-processing procedure by correctly incorporating the winding of charges around fluxes (or vice versa). This has been done with large clusters, e.g., using Entings finite-lattice method \cite{Enting1977}, in order to define particle strings in a fixed gauge-invariant way so that the particle as well as the relevant fluctuations are both present on each cluster. 

However, linked-cluster expansions are often implemented most efficiently by applying a full graph-decomposition so that calculations on individual graphs cost minimal memory and time. In this work we describe how to integrate systematically the non-local effects of the anyonic statistics in a hypergraph decomposition. The latter have recently been established to naturally allow the treatment of multi-site perturbations within linked-cluster expansions \cite{Muehlhauser2022}. Interestingly, they are also suited to set up a linked-cluster expansion to determine the energy of single charge and flux excitations for the topological phase of Kitaev's toric code in the presence of a general uniform field. Specifically, we determine the ground-state energy and the elementary excitation energies of charges and fluxes in the perturbed topological phase.

The paper is organized as follows. In Sec.~\ref{sect::toric-code-field} we introduce the toric code in a field including an extensive introduction to the bare toric code. In Sec.~\ref{sect::toric-code-field_se} we present all generic aspects for series expansions of the perturbed topological phase in the toric code at finite magnetic fields while Sec.~\ref{sect::lce_tc} contains all information for a linked-cluster expansion with full hypergraph decomposition. Final conclusions are drawn in Sec.~\ref{sect::conclusions}.

\section{Kitaev's toric code in a field}
\label{sect::toric-code-field}

We investigate the perturbed topological phase of the toric code in a uniform field. The Hamiltonian of this model is therefore the sum of the toric code $H^{\text{TC}}$ and a general uniform magnetic field
\begin{equation}
\label{eq::tcf}
H^{\rm TCF} =   H^{\text{TC}}  -  \sum_i \vec{h} \cdot \vec{\sigma}_i,  
\end{equation}
where $\vec{h} = (h_x,h_y,h_z)^T \in \mathds{R}^3$ and $\vec{\sigma} = (\sigma^x, \sigma^y, \sigma^z)^T$. In contrast to the bare toric code described in Subsec.~\ref{sect::toric-code}, this model is not exactly solvable and displays a rich quantum phase diagram \cite{Dusuel2011}. For the special case of a magnetic field pointing in $x$- or $z$-direction the model is isospectral to the well-known transverse field Ising model on the dual square lattice in the relevant low-energy sector \cite{Trebst2007, Hamma2008} displaying a second-order quantum phase transition in the 3D Ising universality class. The quantum phase diagram of the toric code in a magnetic field in the $xz$-plane is obtained with series expansion methods \cite{Vidal2009}, quantum Monte Carlo simulations \cite{Tupitsyn2010,Wu2012}, and tensor networks \cite{Iqbal2014}. Notably, the universality class of the quantum phase transition remains 3D Ising except for the symmetric case $h_x=h_z$ where a multicritical point is expected \cite{Vidal2009,Tupitsyn2010,Dusuel2011}. In contrast, for a magnetic field pointing in transverse $y$-direction, the model is dual to the Xu-Moore model \cite{Xu2004} as well as to the quantum compass model \cite{Nussinov2005} and the nature of the phase transition is strongly first order \cite{Vidal2011}. The full extent of the topological phase in the presence of a general uniform magnetic field has been determined by combining high-order series expansions and tensor network algorithms \cite{Dusuel2011} displaying rich physical behavior depending on the magnetic field direction. This includes planes of first- and second-order quantum phase transitions as well as multicritical lines.

In this work we focus on the perturbed topological phase at finite fields and we describe how to set up high-order series expansions. Consequently, we describe the non-trivial unperturbed limit $\vec{h}=0$ of the bare toric code next. 

\subsection{Kitaev's toric code}
\label{sect::toric-code}

\begin{figure}[H]
	\begin{center}
	\includegraphics[]{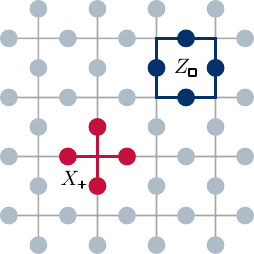} \hspace{2cm} \includegraphics[]{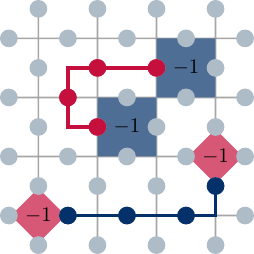}
	\end{center}
	\caption{{\textit Left}: The spin-$1/2$ degrees of freedom are depicted in light blue at the edges of the square lattice. The operators $X_\tcstar$ and $Z_{\tcplaq}$ are illustrated in red or blue respectively. Blue (red) circles indicate that the illustrated operator acts with $\sigma^z$ ($\sigma^x$) on the respective spin. {\textit Right}: Two examples for string operators of the type $S_x$ ($S_z$) in red (blue). Actions of $\sigma^x$ ($\sigma^z$) on spin degrees of freedom are represented by red (blue) circles. The flux (charge) excitations at the end of the string operator $S_x$ ($S_z$) are illustrated as blue (red) squares labeled with $-1$ indicating the negative eigenvalue $z_\tcplaq$ ($x_\tcstar$) of $Z_\tcplaq$ ($X_\tcstar$).
	\label{fig::operators} 
	}
\end{figure}

Kitaev's toric code is defined on a square lattice with spin-$1/2$ degrees of freedom on the edges of the lattice \cite{Kitaev2003}. The Hamiltonian is given by 
\begin{equation} H^{\text{TC}} = - \frac{1}{2}\sum_\tcstar X_\tcstar - \frac{1}{2} \sum_\tcplaq Z_\tcplaq. \end{equation} 
The star operators $X_\tcstar$ and the plaquette operators $Z_\tcplaq$ are both defined as products of four Pauli matrices

\begin{equation}
	X_\tcstar = \prod_{i \in \tcstar} \sigma^x_i, \hspace{2 cm}
	Z_\tcplaq = \prod_{i \in \tcplaq} \sigma^z_i,
\end{equation}
where $\sigma^x_i, \sigma^z_i$ are the usual Pauli matrices. The first product runs over the  four sites surrounding a vertex, the second one over the four sites at the edges of a plaquette as illustrated in Fig.~\ref{fig::operators}. 
As these operators are products of Pauli matrices their eigenvalues $x_\tcstar, z_\tcplaq$ are equal to $\pm1$ \cite{Kitaev2003}.  Furthermore, all these operators commute mutually
\begin{equation}
	\begin{aligned}
	  [X_\tcstar, X_{\tcstar^\prime}] &= 0 \quad \forall \bigtcstar , \bigtcstar^\prime, \\
				 [X_\tcstar, Z_{\tcplaq}] &= 0   \quad \forall \bigtcstar,\bigtcplaq \; , \\
			 [Z_\tcplaq, Z_{\tcplaq^\prime}] &= 0 \quad \forall \bigtcplaq, \bigtcplaq^\prime \;.
			 \end{aligned}
\end{equation}
As a consequence, one can construct a state where all eigenvalues $x_\tcstar, z_\tcplaq$ are equal to $+1$, which corresponds to a ground state
\begin{equation}
	\ket{\rm GS} = \mathcal{N}\prod_\tcstar (1+X_\tcstar) \prod_\tcplaq (1 + Z_\tcplaq) \ket{\text{ref}},
\end{equation}
where $\mathcal{N}$ is a normalization factor and $\ket{\text{ref}}$ is a reference state. 
This reference state has to be chosen such that it is not orthogonal to the ground-state space. For example, the states $\ket{\Uparrow}$ or $\ket{\Rightarrow}$, where all spins point in positive $z$- or $x$-direction can be used. The ground state $\ket{\rm GS}$ is unique on an open plane. However, the ground-state manifold features a non-trivial topological degeneracy which equals $4^g$ on a compact orientable surface of genus $g$ indicating topological order \cite{Kitaev2003}.
Importantly, products of operators $X_\tcstar$ and $Z_\tcplaq$ act trivially on the ground state. Furthermore, any contractible loop of $\sigma^x$ or $\sigma^z$ matrices is equal to the product of operators $X_\tcstar$ or $Z_\tcplaq$ contained in the loop respectively \cite{Kitaev2003}. The ground-state energy is given by $E_0^{\rm tc}=-N/2$ with $N$ the number of spins which equals the total number of plaquettes and stars.

Acting with $\sigma^x_i$ ($\sigma^z_i$) on site $i$ of the ground state creates two flux (charge) excitations corresponding to $z_\tcplaq = -1$ ($x_\tcstar = -1$) adjacent to site $i$. These topological excitations are located at the plaquettes (vertices) of the lattice and behave like hardcore bosons with a mutual anyonic statistics \cite{Kitaev2003, Dusuel2011, Kamfor2014}. More generally, acting with string operators
\begin{equation}
S_z = \prod_{i \in p} \sigma_i^z, \qquad S_x = \prod_{i \in \bar{p}} \sigma^x_i,
\end{equation}
on the ground state,
where $p$ ($\bar{p}$) are open paths on the (dual) lattice, creates two excitations at the end of the respective path \cite{Kitaev2003}. Examples for two such operators are illustrated in the right panel of Fig.~\ref{fig::operators}. Note that it is not possible to have an odd number of excitations on the torus\cite{Kitaev2003}.

 \begin{figure}[ht]
	\begin{center}
	\includegraphics[]{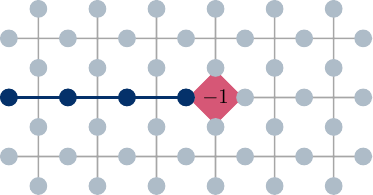} \hspace{1 cm} \includegraphics{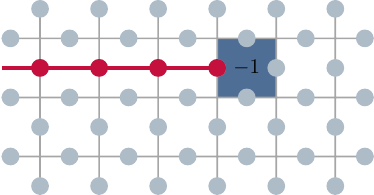}
	\end{center}
	\caption{Canonical one-particle states. {\it Left}: A single charge excitation is obtained by creating two charges from the vacuum and moving one to infinity. This procedure results in a semi-infinite string attached to the charge (indicated in blue). {\it Right}: A single flux excitation is obtained by creating two fluxes from the vacuum and moving one to infinity. This procedure results in a semi-infinite string attached to the flux (indicated in red).
\label{fig::1qp-states}
	}
\end{figure}

However, on an open plane, one can create a pair of excitations and move one of them to infinity. A single charge $x_\tcstar = -1$ at position $\vec{r}$ is denoted by the state $\ket{\vec{r}, \bigtcstar}$. This state is uniquely defined up to the concrete operator which creates the excitation from the ground state. In order to define such an excited state unambiguously we define it by a sequence of Pauli-operators acting on the reference ground state $\ket{\rm GS}$
	\begin{equation}
	\ket{\vec{r}, \bigtcstar} =  \prod_{i \in {p_{\vec{r}}}} \sigma^z_i \ket{\rm GS}, 
	\end{equation}
where $p_{\vec{r}}$ is a straight open path which goes from the excitation to infinity in negative $x$-direction as illustrated in the left panel of Fig.~\ref{fig::1qp-states}. This way we fixed the gauge freedom to define these states uniquely. In the following we call these states \textit{canonical one-charge states}.
In the same fashion one can define \textit{canonical one-flux states}
	\begin{equation}
	\ket{\vec{r}, \bigtcplaq} =  \prod_{i \in {\bar{p}_{\vec{r}}}} \sigma^x_i \ket{\rm GS},
	\end{equation}
where $\bar{p}_{\vec{r}}$ is a straight open path on the dual lattice which is going straight into negative $x$-direction from the position $\vec{r}$ of the flux excitation to infinity as illustrated in the right panel of Fig.~\ref{fig::1qp-states}.
\begin{figure}[ht]
	\begin{center}
		\includegraphics{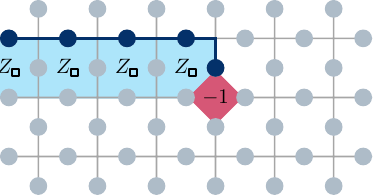} \hspace{1cm} \includegraphics{CanonicalCharge.pdf}
	\end{center}
	\caption{Left figure illustrates a non-canonical one-charge state. The product of a contractible loop operator of $\sigma^z$ with a non-canonical string operator (left) is equal to a canonical string operator (right). Recall that a contractible loop of $\sigma^z$ is equivalent to a product of operators $Z_\tcplaq$.\label{fig::canonicalize_charge}}
\end{figure}
We stress that non-canonical one-particle states with a different string can always be transformed into canonical states by exploiting 

\begin{equation}
	\prod_{\tcplaq} Z_\tcplaq \ket{\rm GS} = \prod_{\tcstar} X_\tcstar \ket{\rm {GS}} = \ket{\rm GS}.
\end{equation}

For a non-canonical one-charge state which is obtained by acting with $\sigma^z_i$ along an open path $l_{\vec{r}}\neq p_{\vec{r}}$ from infinity to $\vec{r}$ this can be achieved in the following way
\begin{equation}
	\prod_{i \in l_{\vec{r}}} \sigma^z_i \ket{\rm GS} = \prod_{i \in l_{\vec{r}}} \sigma^z_i \prod_{\tcplaq \in \Gamma_z} Z_\tcplaq \ket{\rm GS} = \prod_{i \in p_{\vec{r}}} \sigma^z_i \ket{\rm GS} = \ket{\vec{r}, \tcstar},
\end{equation} 
where $\Gamma_z$ is chosen such that
\begin{equation}
 \prod_{i \in l_{\vec{r}}} \sigma^z_i \prod_{\tcplaq \in \Gamma_z} Z_\tcplaq = \prod_{i \in p_{\vec{r}}} \sigma^z_i.
\end{equation} 
One specific example is illustrated for the case of a single-charge state in
Fig.~\ref{fig::canonicalize_charge}. The analogue procedure can be done for corresponding single-flux states. 

More generally, applying the same procedure and conventions, one can generate multi-particle states with an arbitrary number of charges and fluxes. This can be done by successively acting with the corresponding (canonical) string operators on the ground state $\ket{\rm GS}$.

Next we introduce a specific representation of these canonical multi-particle basis states which is convenient for high-order series expansions discussed in the following sections. To this end we consider as reference states $\ket{\text{ref}}$ the fully polarized states $\ket{\Uparrow}$ and $\ket{\Rightarrow}$ in $z$- and $x$-direction, respectively. 
Once the reference state is chosen, all canonical states are uniquely defined including the overall phase. Let us consider an arbitrary multi-particle state with energy 
\begin{equation}
E^{\rm tc}(\{x_\tcstar \}, \{z_\tcplaq \})=-\frac{1}{2}\sum_\tcstar x_\tcstar - \frac{1}{2}\sum_\tcplaq z_\tcplaq 
\end{equation} 
only depending on the eigenvalues of plaquettes and stars. In our construction these states can be written as the product of $\prod_\tcstar (1 \pm X_\tcstar) \prod_\tcplaq (1 \pm Z_\tcplaq)$ and the action of string operators (for each particle one) on the reference state $\ket{\text{ref}}$. The second part of this product corresponds to a spin product state which we call spin background. It can be described by the set of spin eigenvalues $\{s_i\}$ with $s_i\in\{\pm 1\}$ on all sites $i$. For the specific case of a two-particle state with one charge and one flux, one has
\begin{eqnarray}
\ket{\vec{r}_{1,\tcstar },\vec{r}_{2,\tcplaq} } &=&  \left(\prod_{i \in {p_{\vec{r}_1}}} \sigma^z_i \right) \left(\prod_{i \in {\bar{p}_{\vec{r}_2}}} \sigma^x_i\right)  \ket{\rm GS}, \nonumber\\
&=& \mathcal{N}\left( \prod_{\tcstar\neq \tcstar_1} (1+ X_\tcstar)(1- X_{\tcstar_{1}}) \prod_{\tcplaq\neq \tcplaq_{2}} (1 +  Z_\tcplaq)(1- Z_{\tcplaq_{2}})\right)  \left(\prod_{i \in {p_{\vec{r}_1}}} \sigma^z_i \prod_{i \in {\bar{p}_{\vec{r}_2}}} \sigma^x_i \ket{\text{ref}}\right) \nonumber\\
&\equiv & \ket{ \{x_\tcstar \}, \{z_\tcplaq \},  \{s_i\}}
\end{eqnarray}
where $\tcstar_1$ is the star at position $\vec{r}_1$, $\tcplaq_2$ is the plaquette located at $\vec{r}_2$, and $\prod_{i \in {p_{\vec{r}_1}}} \sigma^z_i \prod_{i \in {\bar{p}_{\vec{r}_2}}} \sigma^x_i \ket{\text{ref}}$ is the spin background. 
In general, we keep track of the three sets $\{x_\tcstar \}$, $\{z_\tcplaq \}$, and $\{s_i\}$ of $\mathbb{Z}_2$ degrees of freedom. This way we over-parametrize the states. However, the spin background will turn out to be useful in the next sections to treat the non-trivial statistics of charges and fluxes within the high-order series expansions. 

\section{Series expansions in the perturbed topological phase}
\label{sect::toric-code-field_se}

Our goal is to describe a systematic way how to incorporate non-local anyonic statistics into a high-order series expansion exploiting a full graph decomposition considering the topological phase of the toric code as the unperturbed starting point. How high-order series expansions for arbitrary field directions can be derived using finite clusters is explained in \cite{KamforDiss}, where also Entings finite lattice method \cite{Enting1977, Dusuel2010} has been applied. Such series expansions have been applied successfully to the toric code in a field \cite{Vidal2009,Vidal2011,Dusuel2011}. However, to the best of our knowledge no full graph decomposition has been performed for general field directions. In this section we review how to calculate the ground-state energy and one-quasiparticle excitation energies in the topological phase using perturbation theory about the low-field limit.
To this end we consider perturbative calculations on finite clusters which are sufficiently large to contain the relevant physical properties in the thermodynamic limit \cite{Knetter2003, Dusuel2010, Schulz2012}. The discussion of the full graph decomposition of the perturbed topological phase is then presented in Sec.~\ref{sect::lce_tc}.

\subsection{Set up}
\label{subsect::set_up}
We perform perturbative calculations about the limit $\vec{h}=0$. We therefore take the canonical states $\ket{ \{x_\tcstar \}, \{z_\tcplaq \},  \{s_i\}}$ with energy $E^{\rm tc}(\{x_\tcstar \}, \{z_\tcplaq \})$ as unperturbed basis (see Sec.~\ref{sect::toric-code} for details). We stress again that the unperturbed energies of these basis states do not depend on the spin background $\{s_i\}$. The perturbation $V\equiv -  \sum_i \vec{h} \cdot \vec{\sigma}_i$ is a sum over local terms acting on individual sites $i$. Locally, besides the obvious action on the spin at site $i$, the Pauli matrix $\sigma_i^{x}$ ($\sigma_i^{z}$) flips the two eigenvalues $z_\tcplaq$ ($x_\tcstar$) attached to site $i$ while the Pauli matrix $\sigma_i^{y}$ flips all four eigenvalues of star and plaquette operators containing site $i$. We can therefore split the perturbation $V$ as follows
\begin{equation}
V = T_0 + T_{-2} + T_{+2} + T_{-4} + T_{+4}
\label{eq::decomposition}
\end{equation}
so that $[H^{\rm TC}, T_n] = n T_n$ and $n$ corresponding to the net change of the total charge and flux particle number due to the action of $T_n$. Consequently, at finite fields, the system becomes a challenging quantum many-body problem where the bare number of charges and fluxes is not conserved and the elementary charge and flux excitations gain a finite dispersion and interact.  

The properties of the unperturbed toric code as well as the decomposition Eq.~\ref{eq::decomposition} meet all the requirements to apply the method of perturbative continuous transformation (pCUTs) \cite{Knetter2000, Knetter2003}, which has been done in a series of works \cite{Vidal2009,Vidal2011,Dusuel2011}. The pCUT method allows to map \eqref{eq::tcf} to an effective quasi-particle-number conserving Hamiltonian $H_{\rm eff}$ perturbatively exact up to the calculated order in the parameters $h_x, h_y$, and $h_z$. This effective Hamiltonian obeys $[H^{\rm TC},H_{\rm eff}]=0$, i.e., the effective Hamiltonian is block-diagonal and the quantum many-body problem reduces to a few-body problem in terms of dressed charge and flux excitations. Specifically, $H_{\rm eff}$ can be written as follows
\begin{equation}
 H_{\rm eff} = H^{\rm TC} + \sum_{k\equiv k_x+k_y+k_z=0}^{\infty} h_x^{
 	\vphantom{k_y}k_x} h_y^{k_y} h_z^{\vphantom{k_y}k_z} \sum_{|{\bf m}|=k, M({\bf m})=0} C({\bf m}) T({\bf m})
\label{eq::heff_pcut}
\end{equation}
where ${\bf m}\equiv(m_1,\ldots,m_k)$ with $m_i\in\{0,\pm 2,\pm 4\}$, $T({\bf m}) = T_{m_1} T_{m_2} T_{m_3} \ldots T_{m_k}$ is a product of $k$ operators $T_n$ in order $k$ perturbation theory, $M({\bf m})=\sum m_i$, and $C({\bf m})$ rational numbers. In each perturbative order $k$ one therefore has a weighted sum of quantum fluctuations described by the perturbative process $T({\bf m})$. 

The Hamiltonian \eqref{eq::heff_pcut} is not normal ordered so that physical properties can not be extracted directly. The normal ordering is most efficiently done by calculating matrix elements on finite clusters as only linked processes can have a non-vanishing contribution to these matrix elements. Indeed, if a finite cluster is sufficiently large so that all linked processes of a given perturbative order fit on this cluster for a specific matrix element, then this matrix element can be calculated and is directly valid in the thermodynamic limit. This therefore allows to obtain the normal-ordered form of the effective Hamiltonian.

\begin{figure}
	\begin{center}
		\includegraphics[]{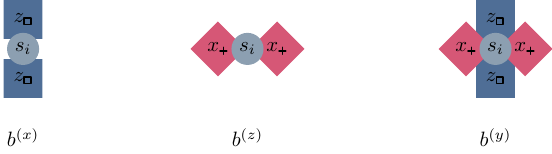}
	\end{center}
	\caption{An illustration of the three different bond types $b^{(\alpha)}_i$ with $\alpha\in\{x,y,z\}$. 
	These bonds contain the respective charge and flux sites as well as the spin $s_i$ which are affected by the action of Pauli matrices $\sigma^x_i, \sigma^z_i$ and $\sigma^y_i$.\label{fig::bondtypes1}}
\end{figure}

The appearance of linked processes can be directly seen by introducing $T_n=\sum_{b^{(\alpha)}}\tau_{n,b^{(\alpha)}}$ where $\tau_{n,b^{(\alpha)}}$ acts on bonds $b^{(\alpha)}$ with $\alpha\in\{x,y,z\}$. These different bond types are illustrated in Fig. \ref{fig::bondtypes1}.
The bond $b^{(\alpha)}$ is the ordered set of charge sites $\bigtcstar$, flux sites $\bigtcplaq$, and spin sites $i$ on which the operator $\tau_{n,b^{(\alpha)}}$ acts. Clearly, any bond contains the spin site $i$ on which the respective Pauli matrix acts. Additionally, all charge and flux sites hosting operators whose eigenvalues $x_\tcstar, z_\tcplaq $ are affected by the respective Pauli matrix need to be included in the bond. Note that it is important to distinguish the different types of sites within the bonds. Here it is necessary to distinguish spin sites from flux and charge sites, as the operators act differently on them. 
Accordingly, a bond $b^{(z)}$ ($b^{(x)}$) includes a spin site and the neighboring charge sites (flux sites), whereas a bond $b^{(y)}$ contains a spin site and the four neighboring charge and flux sites. The effective Hamiltonian can then be exactly rewritten as
\begin{equation}
	H_{\rm eff}=H^{\rm TC} + \sum_{k\equiv k_x+k_y+k_z=0}^{\infty} h_x^{\vphantom{k_y}k_x} h_y^{k_y} h_z^{\vphantom{k_y}k_z}\sum_{|{\bf m}|=k, M({\bf m})=0} C({\bf m}) \sum_{b_1^{(\alpha_1)},b_2^{(\alpha_2)},...b_k^{(\alpha_k)}} \tau_{m_1,b_1^{(\alpha_1)}}\tau_{m_2,b_2^{(\alpha_2)}}...\tau_{m_k,b_k^{(\alpha_k)}}\,.
	\label{eq:pCUT_with_taus}
\end{equation}

In order to determine prefactors of the normal-ordered quasi-particle conserving operators in $H_{\rm eff}$ via the calculation of matrix elements, one has to evaluate $$\bra{\Psi} \tau_{m_1,b_1^{(\alpha_1)}}\tau_{m_2,b_2^{(\alpha_2)}}...\tau_{m_k,b_k^{(\alpha_k)}}\ket{\Phi}\,,$$ where $\ket{\Phi}$ and $\ket{\Psi}$ are states on sufficiently large clusters with the same number of quasi-particles. One therefore has to specify the action of the local operators $\tau_{n,b^{(\alpha)}}$ on the canonical states \mbox{$\ket{ \{x_\tcstar \}, \{z_\tcplaq \},  \{s_i\}}$}. 
We stress that the action on the spin background depends on the choice of the specific reference state $\ket{\text{ref}}$. As an example, we list in Tab.~\ref{table::toperators} the different processes taking the fully polarized state $\ket{\Rightarrow}$ as the reference state. 

As outlined in Sec.~\ref{sect::toric-code-field}, for the perturbed topological phase of the toric code in a field, one has to take care that scalar products are made with canonical states. States arising from the action of the operators $\tau_{n,b^{(\alpha)}}$ are not necessarily canonical. In the following subsections we discuss how to evaluate these scalar products properly for the ground-state energy and for one-quasi-particle states on a single large cluster. In principle the scalar products for higher quasi-particle states are based on the same principles and are described in the literature \cite{KamforDiss}.

\subsection{Ground-state energy}
\label{subsect::groundstate}
Let us start the discussion for the ground-state energy $E_0$ by focusing on a single term of the effective Hamiltonian $H^{\rm eff}$ in order $k$ perturbation theory. Ignoring prefactors in Eq.\eqref{eq::heff_pcut}, this demands to calculate the following expectation value  
\begin{equation} 
 \bra{\rm{GS}} \tau_{m_1,b_1^{(\alpha_1)}}\tau_{m_2,b_2^{(\alpha_2)}}...\tau_{m_k,b_k^{(\alpha_k)}} \ket{\rm GS}\,, \label{eq::gs_matrix_element}
\end{equation}
where $\ket{\rm GS}$ represents the unique ground state of the toric code on the chosen open cluster. 

\begin{table}[H]
	\begin{center}
\begin{tabular}{ | c | c |}
	\hline 
		$\sigma_i^x =\tau_{2,b^{(x)}}+\tau_{0,b^{(x)}}+\tau_{-2,b^{(x)}}$ & $\sigma_i^z=\tau_{2,b^{(z)}}+\tau_{0,b^{(z)}}+\tau_{-2,b^{(z)}}$ \\
	\hline
	$\ket{0 0; 0} \quad \rightarrow\quad \phantom{-}\ket{1 1; 0}$&$\ket{0 0; 0} \quad \rightarrow\quad \phantom{-}\ket{1 1; 1}$\\
	$\ket{0 0; 1} \quad\rightarrow\quad -\ket{1 1; 1}$&$\ket{0 0; 1} \quad\rightarrow\quad \phantom{-}\ket{1 1; 0}$\\
	\hline
	$\ket{0 1; 0} \quad \rightarrow\quad \phantom{-}\ket{1 0; 0}$&$\ket{0 1; 0} \quad \rightarrow\quad \phantom{-}\ket{1 0; 1}$\\
	$\ket{0 1; 1} \quad\rightarrow\quad -\ket{1 0; 1}$&$\ket{0 1; 1} \quad\rightarrow\quad \phantom{-}\ket{1 0; 0}$\\
	$\ket{1 0; 0} \quad\rightarrow\quad \phantom{-}\ket{0 1; 0}$ &$\ket{1 0; 0} \quad\rightarrow\quad \phantom{-}\ket{0 1; 1}$\\
	$\ket{1 0; 1} \quad\rightarrow \quad-\ket{0 1; 1}$&$\ket{1 0; 1} \quad\rightarrow\quad \phantom{-}\ket{0 1; 0}$\\
	\hline
	$\ket{1 1; 1} \quad\rightarrow\quad -\ket{0 0; 1}$&$\ket{1 1; 1} \quad\rightarrow\quad \phantom{-}\ket{0 0; 0}$\\
	$\ket{1 1; 0}\quad \rightarrow\quad \phantom{-}\ket{0 0; 0}$ &$\ket{1 1; 0}\quad \rightarrow\quad \phantom{-}\ket{0 0; 1}$\\
	\hline
\end{tabular}
\end{center}
\caption{On the left the action of $\sigma^x_i$ on the local state configuration $\ket{\tilde{z}_{\tcplaq_1}, \tilde{z}_{\tcplaq_2} ; \tilde{s}_i}$ on a $b^{(x)}$-bond, which contains two plaquette operators and the background spin $s_i$. On the right the action of $\sigma^z$ on the local state $\ket{ \tilde{x}_{\tcstar_1}, \tilde{x}_{\tcstar_2} ; s_i}$ on a $b^{(z)}$-bond type is given, which contains two star operators and the same background spin. The action on the spin background variables $s_i$ depends on the reference states $\ket{\text{ref}}$. Here we have used the fully polarized states $\ket{\Rightarrow}$ in $x$-direction. The upper, middle, and lower line refers to the processes of $\tau_{2,b^{(\alpha)}}$, $\tau_{0,b^{(\alpha)}}$, and $\tau_{-2,b^{(\alpha)}}$ with $\alpha\in\{x,z\}$, respectively. For the sake of clarity we use Boolean values $\tilde{z}_\tcplaq = \tfrac{1}{2}(1-z_\tcplaq)$ and $\tilde{x}_\tcstar = \tfrac{1}{2}(1-x_\tcstar)$, which count the number of excitations at a given $\bigtcstar$ or $\bigtcplaq$ and for the background spins $\tilde{s}_i = \tfrac{1}{2}(1+s_i)$. Furthermore, $\sigma^y=\mathrm{i} \sigma^x \sigma^z$ acts on the background spin $s_i$ as well as on the four eigenvalues of the star and plaquette operators which surround site $i$. \label{table::toperators}
}
\end{table}

The successive action of the $\tau$-operators in Eq.~\eqref{eq::gs_matrix_element} is properly treated on the state level by the processes listed in Tab.~\ref{table::toperators}. Because the product of $\tau$-operators is quasi-particle conserving, it can not create any excitations. Instead, the final state $\ket{\rm f}\equiv\tau_{m_1,b_1^{(\alpha_1)}}\tau_{m_2,b_2^{(\alpha_2)}}\ldots\tau_{m_k,b_k^{(\alpha_k)}} \ket{\rm GS}$ is either zero or is the ground state $\ket{\rm GS}$ with an additional phase factor $\left[ i^{n_a} \right] $ with $n_a \in \{0,1,2,3\}$ resulting from the action of the $\tau$-operators.

One can explicitly write down the final state $\ket{\rm f}$ as  
\begin{equation}\label{eq:sigmas}
	 \ket{\rm f} = \mathcal{N}\prod_\tcstar (1+X_\tcstar) \prod_\tcplaq (1 + Z_\tcplaq)\;\sigma^{\alpha_1}_{i_1}\sigma^{\alpha_2}_{i_2}\ldots\sigma^{\alpha_k}_{i_k}\ket{\rm ref}\, .
\end{equation}
The non-contributing sequences are easily identified, so we restrict the discussion to the contributing ones.
For the explicit reference state $\ket{\Rightarrow}$, the action of $\sigma^{\alpha_1}_{i_1}\sigma^{\alpha_2}_{i_2}\ldots\sigma^{\alpha_k}_{i_k}$ on the reference state is given as
\begin{equation}
	\label{eq:vacuum-loops-Z}
\sigma^{\alpha_1}_{i_1}\sigma^{\alpha_2}_{i_2}\ldots\sigma^{\alpha_k}_{i_k} \ket{\Rightarrow}
= \left[\mathrm{i}^{n_a} \right] \ket{s^x} = \left[\mathrm{i}^{n_a}\right] \prod_{\tcplaq \in \Gamma_z} Z_\tcplaq \ket{\Rightarrow}\,,
\end{equation}
 $\ket{s^x}$ is a product state in the $\sigma^x$-basis and $\left[\mathrm{i}^{n_a} \right]$ with $n_a \in \{0,1,2,3\}$ is the phase factor resulting from the action of the Pauli matrices, whereas $\Gamma_z$ is a set of plaquettes such that 
 \begin{equation}	
 		\ket{s^x} = \prod_{\tcplaq \in \Gamma_z} Z_\tcplaq \ket{\Rightarrow}.
 \end{equation}
 Note that we are in the ground-state sector, so $\ket{s^x}$ can at most contain several contractible loops of flipped spins in the spin-background, which can be replaced by a product of operators $Z_\tcplaq$.
Inserting Eq.~\eqref{eq:vacuum-loops-Z} into Eq.~\eqref{eq:sigmas} yields 
\begin{equation} \label{eq:final-gs-right}
	\left[\mathrm{i}^{n_a}\right] \cdot \mathcal{N}  \prod_{\tcplaq \in \Gamma_z} Z_\tcplaq \prod_\tcstar (1+X_\tcstar) \prod_\tcplaq (1 + Z_\tcplaq) \ket{\Rightarrow} =\left[\mathrm{i}^{n_a}\right]  \prod_{\tcplaq \in \Gamma_z} Z_\tcplaq \ket{\rm GS} =\left[\mathrm{i}^{n_a}\right] \ket{\rm GS}.
\end{equation}
Using the operators defined in Tab. \ref{table::toperators} results exactly in the middle part of Eq. \eqref{eq:final-gs-right}, i.e. they yield the result in terms of the state $ \prod_{\tcplaq \in \Gamma_z}Z_\tcplaq \ket{\rm GS}$  and correctly include the additional phase $\left[\mathrm{i}^{n_a}\right]$ within the amplitude of this state.

Interestingly, the state $\ket{\rm f}$ can also be evaluated for an arbitrary reference state. To this end we sort the product of Pauli matrices in Eq.~\eqref{eq:sigmas} by their flavor taking into account the non-trivial commutation relations. Note that Pauli matrices $\sigma^y$ can always be written as ${\rm i}\sigma^x\sigma^z$. In the ground-state sector one can always write the obtained sorted expression as the product of contractible loops of $\sigma^z$ or $\sigma^x$.
 Each contractible loop operator can be replaced exactly by the product of plaquette or star operators contained in the respective loop
 \begin{equation}
 \sigma^{\alpha_1}_{i_1}\sigma^{\alpha_2}_{i_2}\ldots\sigma^{\alpha_k}_{i_k} =   \left[\mathrm{i}^{n_a}\right]  \prod_{\tcstar \in \Gamma_x} X_{\tcstar} \prod_{\tcplaq \in \Gamma_z} Z_\tcplaq\,.
 \end{equation}
Since any product of star and plaquette operators yields identity on any ground state, one directly gets the canonical ground state $\ket{\rm GS}$ with an additional phase factor $\left[\mathrm{i}^{n_a}\right]$ with $n_a \in \{0,1,2,3\}$, which stems from the sorting of the Pauli matrices
	\begin{equation}
	\left[\mathrm{i}^{n_a}\right] \cdot \mathcal{N} \prod_{\tcstar \in \Gamma_x} X_{\tcstar} \prod_{\tcplaq = \Gamma_z} Z_\tcplaq \prod_\tcstar (1+X_\tcstar) \prod_\tcplaq(1 + Z_\tcplaq) \ket{\rm ref} =  \left[\mathrm{i}^{n_a} \right] \ket{\rm GS}\, .
	\end{equation}
So once the phase factor is determined, no post-processing is necessary in the ground-state sector and one has $\bra{\rm{GS}} \tau_{m_1,b_1^{(\alpha_1)}}\tau_{m_2,b_2^{(\alpha_2)}}...\tau_{m_k,b_k^{(\alpha_k)}} \ket{\rm GS}=\left[\mathrm{i}^{n_a}\right]$.

\subsection{One-quasi-particle energies}
\label{subsect::1qp}

The one-quasi-particle sector of the effective Hamiltonian $H_{\rm eff}$ represents two one-particle problems: One for a single charge quasi-particle and one for a single flux quasi-particle. Both problems are exactly decoupled since the magnetic field does not contain any process transforming a charge into a flux or vice versa due to parity conservation of fluxes and charges. Furthermore, there is an exact self-duality in the toric code in a field so that charge and flux excitation energies are identical up to an interchange of $h_x$ and $h_z$.

Let $\ket{\vec{r}_1}$ be either the canonical one-charge state $\ket{\vec{r}_1, \bigtcstar}$ or the canonical one-flux state $\ket{\vec{r}_1, \bigtcplaq}$. In real space one then has to calculate the one-particle hopping amplitudes $a_{\vec{\delta}}$ given by 
\begin{equation} 
 a_{\vec{\delta}} = \bra{\vec{r}_1+\vec{\delta}} H_\text{eff}-E_0 \ket{\vec{r}_1} \,.
	\end{equation}
Introducing corresponding one-quasi-particle states in Fourier space $\ket{\vec{k}}\equiv \sqrt{\frac{2}{N}}\sum_{\vec{r}_1} \exp(\mathrm{i} \vec{k} \vec{\delta})\ket{\vec{r}_1}$ with $N$ the number of spins, the one-particle dispersion of charges and fluxes is 
\begin{equation}\label{eq::disp}
\omega(\vec{k})\equiv \bra{\vec{k}} H_\text{eff}-E_0 \ket{\vec{k}}\,.
\end{equation}
The minimum of this dispersion is called the one-particle gap $\Delta$ located at momentum $\vec{k}=0$ for charges and fluxes. In the following we call the charge gap $\Delta^{\tcstar}$ and the flux gap $\Delta^{\tcplaq}$. 

Next we discuss how to evaluate the matrix elements $a_{\vec{\delta}}$. As in the calculation of the ground-state energy, we can focus on the contribution of a single perturbative term 
	\begin{equation}
		\bra{\vec{r}_2} \tau_{m_1,b_1^{(\alpha_1)}}\tau_{m_2,b_2^{(\alpha_2)}} \ldots\tau_{m_k,b_k^{(\alpha_k)}} \ket{\vec{r}_1}\equiv \langle\vec{r}_2| {\rm f} \rangle
	\end{equation} 	
to the hopping amplitude $a_{\vec{\delta}}$ with $\vec{r}_2\equiv\vec{r}_1+ \vec{\delta}$. In order to evaluate the scalar product $\langle\vec{r}_2| {\rm f} \rangle$, one has to express $\ket{\rm{f}}$ in terms of a canonical one-quasi-particle state along the lines discussed in Sect.~\ref{sect::toric-code-field}.

First let us rewrite a canonical one-charge state as
\begin{equation}
\ket{\vec{r}_1, \bigtcstar}  = \prod_{i \in {p_{\vec{r}_1}}} \sigma^z_i \ket{\rm GS} = 
\mathcal{N} (1- X_{\tcstar_1}) \prod_{\tcstar \neq \tcstar_1} (1+X_\tcstar)  \prod_\tcplaq(1+Z_\tcplaq) \prod_{i \in {p_{\vec{r}_1}}} \sigma^z_i \ket{\rm ref}.\\ 
\label{eq::one_charge_general}
\end{equation}
For a given reference state $\ket{\rm ref}$ this defines a canonical spin-background for one-charge states $\prod_{i \in {p_{\vec{r}_1}}} \sigma^z_i \ket{\rm ref}.$
Next we investigate the action of a non-vanishing quasi particle number conserving operator sequence on a canonical one-charge state	
	\begin{equation}
		\begin{aligned}
			\ket{\rm f} &= \tau_{m_1,b_1^{(\alpha_1)}}\tau_{m_2,b_2^{(\alpha_2)}} \ldots \tau_{m_k,b_k^{(\alpha_k)}} \ket{\vec{r}_1, \bigtcstar}  \\  
	& =	\mathcal{N}  (1 - X_{\tcstar_2}) \prod_{\tcstar \neq \tcstar_2} (1+X_{\tcstar})  \prod_\tcplaq(1 + Z_\tcplaq) \; \sigma^{\alpha_1}_{i_1} \sigma^{\alpha_2}_{i_2} \ldots \sigma^{\alpha_k}_{i_k} \prod_{i \in {p_{\vec{r}_1}}} \sigma^z_i \ket{\rm ref}\, ,
	\end{aligned}	
	\end{equation}
where $\bigtcstar_2$ is the charge site at position $\vec{r}_2$. While it is clear that the charge moved from $\vec{r}_1$ to $\vec{r}_2$, the spin-background is not necessarily in a canonical form.

At this point it is most instructive to use $\ket{\Rightarrow}$ as reference state and to bring the spin-background into a canonical form. For this choice we have
	\begin{equation}
	 \ket{\rm{f}} = \mathcal{N} (1 - X_{\tcstar_2})\prod_{\tcstar \neq \vec{\tcstar_2}} (1+X_{\tcstar})   \prod_\tcplaq(1 + Z_\tcplaq)\; \sigma^{\alpha_1}_{i_1} \sigma^{\alpha_2}_{i_2} \ldots \sigma^{\alpha_k}_{i_k} \, \prod_{i \in {p_{\vec{r}_1}}} \sigma^z_i \ket{\Rightarrow}.
	\end{equation}
	Using $\sigma^y = \mathrm{i} \sigma^x \sigma^z$, the action of the Pauli matrices on $\ket{\Rightarrow}$ is evaluated
	\begin{equation}
		\label{eq:phase-factor-a0}
		\sigma^{\alpha_1}_{i_1} \sigma^{\alpha_2}_{i_2} \ldots \sigma^{\alpha_k}_{i_k} \prod_{i \in {p_{\vec{r}_1}}} \sigma^z_i \ket{\Rightarrow} = 
		\left[\mathrm{i}^{n_a}\right]\cdot \ket{s^x} = \left[\mathrm{i}^{n_a}\right]\cdot \prod_{\tcplaq \in \Gamma_z} Z_\tcplaq \prod_{i \in {p_{\vec{r}_2}}}  \sigma^z_i\, \ket{\Rightarrow}\,,
	\end{equation}
where $\ket{s^x}$ is a product state in the $\sigma^x$-basis and the phase factor $\mathrm{i}^{n_a}$ with $n_a \in \{0,1,2,3\}$ results from the action of the Pauli matrices.
 
The expression on the right results from the fact that any $\ket{s^x}$ can be written as a product of $\sigma^z$ matrices acting on the reference state $\ket{\Rightarrow}$. In the current case this product of $\sigma^z$ can always be chosen as the canonical Pauli string operator on $p_{\vec{r}_2}$ of the one-charge state $\ket{\vec{r}_2,\bigtcstar}$ times some product of $Z_\tcplaq $ over the appropriate set $\Gamma_z$. With this the final state $\ket{\rm f}$ becomes
		\begin{equation}
		\begin{aligned}
		\ket{\rm{f}} =& \left[\mathrm{i}^{n_a}\right] \cdot \mathcal{N}(1 - X_{\tcstar_2}) \prod_{\tcstar \neq \tcstar_2} (1+X_{\tcstar})   \prod_\tcplaq(1 + Z_\tcplaq)  \prod_{\tcplaq \in \Gamma} Z_\tcplaq \prod_{i \in {p_{\vec{r}_2}}} \sigma^z_i \ket{\Rightarrow}
		\\
		=& \left[\mathrm{i}^{n_a}\right] \cdot \mathcal{N} (1 - X_{\tcstar_2})\prod_{\tcstar \neq \tcstar_2} (1+X_{\tcstar})   \prod_\tcplaq(1 + Z_\tcplaq)  \prod_{i \in {p_{\vec{r}_2}}} \sigma^z_i \ket{\Rightarrow} \\
		=& \left[\mathrm{i}^{n_a}\right] \ket{\vec{r}_2, \bigtcstar }  \, .
		\end{aligned}
		\label{eq::just-one-phase-factor}
	\end{equation}
Note that the result is again obtained in terms of a canonical state $\ket{\vec{r}_2, \bigtcstar }$  and an additional phase factor $\left[\mathrm{i}^{n_a}\right]$ stemming from the action of the Pauli matrices on the reference state as detailed in Eq. \ref{eq:phase-factor-a0}. The operators given in Tab. \ref{table::toperators} yield the result in terms of a state $\ket{\rm \tilde{f}}$ with a phase $\left[\mathrm{i}^{n_a}\right]$, i.e.,
\begin{equation}
\ket{\rm f } =  \left[\mathrm{i}^{n_a}\right] \ket{\rm \tilde{f}}\,. 
\end{equation}
So although the spin-background of $\ket{\rm\tilde{f}}$ is not necessarily canonical we have
\begin{equation}
	\label{eq:trivial_canonicalization}
\ket{\rm \tilde{f}} = \ket{\vec{r}_2, \bigtcstar}, 
\end{equation}
reflecting that canonicalizing the state $\ket{\rm \tilde{f}}$ is a trivial operation due to the absence of fluxes.

Note that this is different if we choose the reference state $\ket{\Uparrow}$ and act with the Pauli matrices to the right
	\begin{equation}
	\ket{\rm{f}} = \mathcal{N} (1 - X_{\tcstar_2})\prod_{\tcstar \neq \tcstar_2} (1+X_{\tcstar})  \prod_\tcplaq(1 + Z_\tcplaq) \; \sigma^{\alpha_1}_{i_1} \sigma^{\alpha_2}_{i_2} \ldots \sigma^{\alpha_k}_{i_k} \prod_{i \in {p_{\vec{r}_1}}} \sigma^z_i \ket{\Uparrow}\,.
	\end{equation}
	Then we can write
	\begin{equation}
	\sigma^{\alpha_1}_{i_1} \sigma^{\alpha_2}_{i_2} \ldots \sigma^{\alpha_k}_{i_k} \prod_{i \in {p_{\vec{r}_1}}} \sigma^z_i  \ket{\Uparrow} =  \left[\mathrm{i}^{n_a}\right]\cdot \ket{s^z} = \left[\mathrm{i}^{n_a}\right]\cdot \prod_{\tcstar \in \Gamma_x} X_\tcstar \ket{\Uparrow}\, ,
	\end{equation}
	where the factor $[\mathrm{i}^{n_a}]$ comes from the action of the Pauli operators. We can then commute the product over $X_\tcstar$ to the left
	\begin{equation}
	\ket{\rm{f}} = \left[\mathrm{i}^{n_a}\right]\cdot  \left( \prod_{\tcstar \in \Gamma_x} X_\tcstar\right) \, \mathcal{N}   (1 - X_{\tcstar_2}) \prod_{\tcstar \neq \tcstar_2} (1+X_{\tcstar}) \prod_\tcplaq(1 + Z_\tcplaq)  \ket{\Uparrow}, \label{eq::more-than-one-phase-factor}
	\end{equation}
	which is a product of $X_\tcstar$ acting on a one-charge state. And hence we get the following expression 
		\begin{equation}
	\ket{\rm{f}} = \left[\mathrm{i}^{n_a+2 n_b}\right]\cdot \mathcal{N} (1 - X_{\tcstar_2}) \prod_{\tcstar \neq \tcstar_2} (1+X_{\tcstar})  \prod_\tcplaq(1 + Z_\tcplaq)  \ket{\Uparrow},
	\end{equation}
	where 
	\begin{equation}
		n_b = \begin{cases}
		1 \quad \text{if} \quad \bigtcstar_2 \in \Gamma_x\\
		0 \quad \text{else}.
		\end{cases}
	\end{equation}
	Clearly, in this convention, the definition of the operators in Tab.~\ref{table::toperators} needs to be adapted. These adapted operators then also yield the result in the form
	\begin{equation}
		\ket{\rm f} = \left[\mathrm{i}^{n_a}\right]	\ket{\rm \tilde{f}}\,.
	\end{equation}
	However, in contrast to Eq.~\eqref{eq:trivial_canonicalization}, we have
	\begin{equation}
		\ket{\rm \tilde{f}} = \left[\mathrm{i}^{2 n_b}\right] \ket{\vec{r}_2, \bigtcstar}\,.
	\end{equation}
	So we still need to determine $n_b$ from the spin-background of the state $\ket{\rm \tilde{f}}$ by checking whether there are contractible loops surrounding the charge at $\vec{r}_2$. Physically, each such loop corresponds to the winding of a flux around the charge. As a consequence, the total effect of all these loops depends only on the parity of the loop number. In case of even (odd) parity  no (an) additional sign results which has to be included in this non-trivial postprocessing procedure.\\
	
	In the previous part we used an explicit reference state because it is easy to act with sequences of Pauli operators on product states in the spin basis. But actually the arguments are independent of the reference state. Note that we effectively just sorted the Pauli operators by flavor.
	Splitting $\sigma^y = \mathrm{i} \sigma^x \sigma^z$ we arranged the product $\sigma^{\alpha_1}_{i_1} \sigma^{\alpha_2}_{i_2} \ldots \sigma^{\alpha_k}_{i_k}$
	such that $\sigma^x_i$ acts before $\sigma^z_i$ or vice versa.
	Accordingly, on an operator level, the signs stem from the ordering of the Pauli operators.
	In this sense the $\tau$-operators implement a bookkeeping technique for the necessary reordering and yield the appropriate sign for the ordered product.
	
As for the ground-state energy, the state $\ket{\rm f} $ can then be written independently of the reference state as
	\begin{equation}
		\ket{\rm f} =	\left[\mathrm{i}^{n_a}\right]\cdot \mathcal{N}(1 - X_{\tcstar_2}) \prod_{\tcstar \neq \tcstar_2} (1+X_{\tcstar})   \prod_\tcplaq(1 + Z_\tcplaq) \varsigma^z \varsigma^x  \ket{\rm ref}\,, 
	\end{equation}
	where $\varsigma^\alpha$ is the part consisting of $\sigma^\alpha$. The additional factor $\left[\mathrm{i}^{n_a}\right]$ stems from the commutation relations of the Pauli matrices. 
	It is clear that $\varsigma^x$ corresponds to contractible loop operators and hence we have
		\begin{equation}
		\ket{\rm f} = \left[\mathrm{i}^{n_a}\right]\cdot \mathcal{N} (1 - X_{\tcstar_2}) \prod_{\tcstar \neq \tcstar_2} (1+X_{\tcstar}) \prod_\tcplaq(1 + Z_\tcplaq)  \varsigma^z \prod_{\tcstar \in \Gamma_x}  X_\tcstar  \ket{\rm ref}. \label{eq::pseudocanonical}
	\end{equation}
	The product $\varsigma^z$ corresponds to a product of stabilizers over the set $\Gamma_z$ with a canonical path going from $\vec{r}_2$ to infinity. One therefore has
	\begin{equation}
		\ket{\rm f} = \left[\mathrm{i}^{n_a}\right]\cdot  \mathcal{N}   (1 - X_{\tcstar_2}) \prod_{\tcstar \neq \tcstar_2} (1+X_{\tcstar}) \prod_\tcplaq(1 + Z_\tcplaq) \prod_{\tcplaq \in \Gamma_z} Z_\tcplaq \prod_{i \in {p_{\vec{r}_2}}} \sigma^z_i \prod_{\tcstar \in \Gamma_x} X_\tcstar  \ket{\rm ref}\, .
	\end{equation}	
As a next step we move the product of $X_\tcstar$ and $Z_\tcstar$ to the left	
	\begin{equation}
		\ket{\rm f} = \left[\mathrm{i}^{n_a+2n_b}\right]\cdot \left(  \prod_{\tcstar \in \Gamma_x}   X_\tcstar\right)\left(  \prod_{\tcplaq \in \Gamma_z} Z_\tcplaq    \right) \mathcal{N} (1 - X_{\tcstar_2})  \prod_{\tcstar \neq \tcstar_2} (1+X_{\tcstar})  \prod_\tcplaq(1 + Z_\tcplaq)  \prod_{i \in {p_{\vec{r}_2}}} \sigma^z_i  \ket{\rm ref}\,,
	\end{equation}
	where $n_b$ captures whether $\prod_{\tcstar \in \Gamma_x}   X_\tcstar$ and $\prod_{i \in {p_{\vec{r}_2}}} \sigma^z_i$ commute or anticommute
	\begin{equation}
		q_2 = \begin{cases} 1 \quad \text{if} \quad 	\bigtcstar_2 \in \Gamma_x
				\\ 0  \quad \text{else}
		\end{cases}
	\end{equation}
	What remains is the action of stabilizer operators on a canonical one-charge state
	\begin{equation}
	\begin{aligned}
	 \ket{\rm f} = \left[\mathrm{i}^{n_a+2n_b}\right]\cdot \left(\prod_{\tcstar \in \Gamma_x} X_\tcstar\right)\left( \prod_{\tcplaq \in \Gamma_z} Z_\tcplaq\right)  \ket{\vec{r}_2, \tcstar} = \left[\mathrm{i}^{n_a+2n_b+2n_c}\right] \ket{ \vec{r}_2, \tcstar},
	\end{aligned}
	\end{equation}
 	where
 	\begin{equation}
 	n_c = n_b = \begin{cases} 1 \quad \text{if} \quad 	\bigtcstar_2 \in \Gamma_x
 		\\ 0  \quad \text{else.}
 	\end{cases}
 	\end{equation}
 	As $n_b \in \{0,1\}$ we have 
	\begin{equation}
		\label{eq:only-na1}
	\begin{aligned}
			\ket{\rm f} = \left[\mathrm{i}^{n_a+2n_b+2n_c}\right] \ket{ \vec{r}_2, \tcstar} = 	\left[\mathrm{i}^{n_a+4n_b}\right]  \ket{ \vec{r}_2, \tcstar}  = 	\left[\mathrm{i}^{n_a}\right]  \ket{ \vec{r}_2, \bigtcstar},
	\end{aligned}
	\end{equation}
	which means that the phase factor stems from the sorting of the Pauli matrices.
	
	Note that we did choose a convention on the sorting of the Pauli matrices in this calculation, by sorting $\sigma^x$ to the right. 
	The operators given in Tab.~\ref{table::toperators} take into account the signs arising from this reordering and yield the final state in the form
	\begin{equation}
		\ket{\rm f} = \left[\mathrm{i}^{n_a}\right] \ket{\rm \tilde{f}}\, . 
	 \end{equation}
	 Comparing to Eq.~\eqref{eq:only-na1}, one sees that the canonicalization of $\ket{\rm \tilde{f}}$ cannot yield an additional phase factor.
	 
	 Overall, these considerations allow to set up series expansions of the perturbed topological phase in the zero- and one-particle sector using sufficiently large clusters. A generalization to many particles requires further considerations treating the spin background properly in the evaluation of scalar products between multi-particle states \cite{KamforDiss,Kamfor2014}. Next we go beyond series expansions on large clusters by performing a full graph decomposition. 
	 
\section{Hypergraph decomposition for the perturbed topological phase}
\label{sect::lce_tc}
In this section we describe how to execute linked-cluster expansions \cite{Gelfand2000, Oitmaa2006} for the perturbed toric code in the topological phase at finite fields using a full hypergraph decomposition \cite{Muehlhauser2022}. In this work we restrict ourselves to the zero- and one-particle sector. In these sectors it turns out that the non-trival mutual statistics can be correctly included by taking the semi-infinite string operators into account within the hypergraph decomposition.

 
\subsection{Linked-cluster expansions}
\label{subsect::lce}

On a sufficiently large cluster $C$ as introduced in Sec.~\ref{sect::toric-code-field_se}, the ground-state energy as well as the irreducible one-particle matrix elements of the effective Hamiltonian can be written as \cite{Gelfand2000}
\begin{equation} K(C) = \kappa(C) + \sum_{c \subset C} \kappa(c), \label{eq::linked_cluster1} \end{equation}
where $\kappa(C)$ is the reduced contribution of the cluster $C$, i.e., the contribution which contains only the perturbative processes which act at least once on every bond of the cluster $C$ \cite{Gelfand2000, Coester2015}. 
Note that the sum runs over all sub-clusters $c$ of $C$ and not over all distinct sub-clusters of $C$. 
For the ground-state energy, the contribution $K(c)$ of a cluster $c$ is 
\begin{equation}
	E_0(c) = \bra{\rm GS} H_{\rm eff}^c \ket{\rm GS},
\end{equation} 
where $H_{\rm eff}^c$ ist the restriction of $H_{\rm eff}$ to the bonds and sites of the cluster $c$. Accordingly, $\ket{\rm GS}$ is the unperturbed ground state on the full system. Obviously, $H_{\rm eff}^c$ can also be evaluated on the restriction of $\ket{\rm GS}$ to the cluster $c$. 
For the one-quasiparticle states, we subtract the ground-state energy and calculate matrix elements of the form \cite{Gelfand1996}
\begin{equation}
	t_{\vec{r}, \vec{r}+\vec{\delta}} = \bra{\vec{r}+\vec{\delta}} H_{\rm eff} - E_0 \ket{\vec{r}}.
\end{equation}
The contribution $K(c)$ of cluster $c$ is then 
\begin{equation}
	t_{\vec{r}, \vec{r}+\vec{\delta}}^c = \bra{\vec{r}+\vec{\delta}} H_{\rm eff}^c - E_0(c) \ket{\vec{r}},
\end{equation}
where $\ket{\vec{r}}$ is a canonical state with a flux or charge at position $\vec{r}$ defined on the full system. 
The reduced contribution $\kappa(C)$ can be calculated by
	\begin{equation} \kappa(C) = K(C) - \sum_{c \subset C} \kappa(c), \end{equation}
where the sum runs over all proper sub-clusters of $c$, which means that the reduced contribution depends only on reduced contributions of smaller clusters and can be determined recursively. The restrictions of the Hamiltonian on isomorphic (structurally equivalent) clusters are equivalent \cite{Gelfand2000} and thus equivalent matrix elements of the (effective) Hamiltonian on these clusters agree.
So we can rewrite \eqref{eq::linked_cluster1} as 
\begin{equation} K(C) = \sum_{\mathcal{E} \subseteq C}   N(\mathcal{E}, C)\cdot \kappa(\mathcal{E})\,,\label{eq::linked_cluster2}\end{equation}
where $\mathcal{E}$ labels an equivalence class of sub-clusters, $N(\mathcal{E}, C)$ gives the number of sub-clusters of $C$ which belong to $\mathcal{E}$, and $\kappa(\mathcal{E})$
is the reduced contribution of a cluster from the class $\mathcal{E}$ \cite{Gelfand2000,Oitmaa2006}. For extensive quantities $N(\mathcal{E}, C)$  is normalized (usually to the number of sites in the cluster $C$). 
Importantly, all clusters $c$ and $c^\prime$ which are in the equivalence class $\mathcal{E}$ fulfill
\begin{equation}
\kappa(c) = \kappa(c^\prime) = \kappa(\mathcal{E})\,. 
\end{equation}
The distinction of the equivalence classes is based on the structure of the clusters and the states on the clusters within the considered matrix element.
To this end the clusters are usually associated to graphs, where the vertices represent sites and bonds linking these sites correspond to edges.
As structural equivalent clusters are represented by isomorphic graphs we refer to them as isomorphic clusters.
Further, the local states on the clusters within the considered matrix element are incorporated using additional vertex colors. The equivalence of clusters then corresponds to (color-preserving) isomorphism of the respective graphs.

However, for the problem at hand the bonds $b^{(x)}, b^{(y)}, b^{(z)}$ join multiple sites, and not necessarily in a symmetric way, so there is also some orientation within the bonds. 
A representation in terms of hypergraphs \cite{Berge1973} appears natural for multi-site couplings. Interestingly, hypergraphs can be unambiguously represented by bipartite graphs, which are referred to as the König representation \cite{Zykov1974, Konstantinova1995Molecular}.   
So instead of the usual graph representation we use an enhanced König representation to represent the clusters by graphs as described in \cite{Muehlhauser2022}. 
Also in this case information about the local states on the clusters are incorporated into the graph representation using additional vertex colors.
We concretely discuss how to construct the graph representation of the clusters including some concrete examples in Sec \ref{subsect::technicalities}.

Furthermore, for a given perturbation order, the sums in Eqs.~\eqref{eq::linked_cluster1} and \eqref{eq::linked_cluster2} need to consider only clusters up to a given number of bonds, because a cluster can not contribute in perturbation orders smaller than its number of bonds \cite{Gelfand2000}. As the pCUT method is cluster-additive, $\kappa(C)$ vanishes for any disconnected cluster $C$ and so it is sufficient if these sums run over all relevant connected sub-clusters or the corresponding equivalence classes respectively.
Typically the cluster $C$ is chosen large enough to host all perturbative contributions in the desired perturbation order, so that the factors $N(\mathcal{E},C)$ on the finite cluster $C$ are the same as on the infinite lattice $\mathcal{L}$ (after appropriate normalization in the case of extensive quantities). 
Additional selection rules \cite{He1990, Oitmaa2006, Coester2015Hypercubic} to check whether a cluster (or an entire equivalence class) can contribute at a given order are important to further truncate the sums and increase the efficiency of the method. We give some details about such heuristics for the toric code in a field in App.~\ref{sect::non_contributing}.
\subsection{Contributions from individual clusters}
\label{subsect::individual_clusters}
For the low-field limit of the toric code in a general uniform field the unperturbed Hamiltonian depends on the eigenvalues $x_\tcstar, z_\tcplaq$ which are defined locally at the charge or flux sites. The general uniform field then acts as a perturbation on three different types of bonds as described in the last section.
These three bonds correspond to the elementary building blocks of the clusters. This means every cluster relevant for a perturbative expansion is determined by a set of bonds, which is a subset of all the bonds in the lattice. Clusters consisting of a single site are relevant only in zeroth order, so their contributions can be easily calculated and are neglected in the following discussion. Note that in contrast to the hypergraph decomposition in Ref. \cite{Muehlhauser2022}, we have several types of sites in the current problem: The spin-sites $s_i$, the flux-sites $\bigtcplaq$, and the charge-sites $\bigtcstar$.

In Sec.~\ref{sect::toric-code-field_se} we describe how to calculate the ground-state energy and one-quasiparticle matrix elements on finite clusters. In principle each of these calculations can be decomposed as given in Eq.~\eqref{eq::linked_cluster1}. However, for a given irreducible matrix element, the contribution of a cluster can depend on particles residing outside of the clusters in the current case, which is counter-intuitive and a consequence of the non-local properties of topological phases.  Such effects are however absent in the ground-state sector. Indeed, we showed that for the ground-state energy it is sufficient to act with the operators $\tau_{n,b^{(\alpha )}}$ (see Tab.~\ref{table::toperators}) to the right so that the result is given in terms of the canonical ground state $\ket{\rm GS}$ times an additional phase factor. So the ground-state energy can easily be evaluated on each individual cluster and does not depend on any degree of freedom which is outside of the respective cluster.

In the one-quasiparticle sector, the procedure depends on the choice of the reference state. First, there are the specific reference states where the strings of the considered quasi-particle are visible but the corresponding strings of the other quasi-particle type are not visible in the spin-background. Let us consider the generic example of a single charge excitation taking the reference state $\ket{\Rightarrow}$. The phase factor $[\mathrm{i}^{n_a}]$ in the one-charge state in Eq.~\eqref{eq::just-one-phase-factor} originates from the action of the local operators $\tau_{n,b^{(\alpha )}}$ on the canonical state. Due to the absence of fluxes and the specific choice of the reference state, this one-charge state is a canonical state up to the phase $[\mathrm{i}^{n_a}]$.
As a consequence, we can evaluate these irreducible one-charge hopping elements on individual clusters without the necessity to take degrees of freedom outside the cluster into account. 

In contrast, using the reference state $\ket{\Uparrow}$ as in Eq.~\eqref{eq::more-than-one-phase-factor}, the phase factor $[\mathrm{i}^{n_a + 2 n_b}]$ stems from the evaluation of the $\tau_{n,b^{(\alpha )}}$ on this reference state and the canonicalization of the resulting state. In this case the state resulting from the action of the $\tau_{n,b^{(\alpha )}}$ may contain loops which surround the charge and hence lead to an additional phase factor of $-1$.
In the formulation for arbitrary reference states, these conventions correspond to different sorting of the product of Pauli matrices acting on the reference state. In the first convention $\sigma^x$ is sorted to the right of $\sigma^z$, in the second convention it is the other way around.
Overall, in the appropriate convention the evaluation of the sequence of $\tau_{n,b^{(\alpha )}}$ yields a state which is equivalent to a canonical state and an additional phase factor. So these irreducible hopping elements are easily evaluated on each cluster and do not depend on any degree of freedom outside of the cluster. Clearly, by the duality of charges and fluxes also one-flux irreducible matrix elements can be determined on the individual clusters, by choosing the appropriate convention.

Note, however, that in order to evaluate the contribution of a cluster we need to take into account all the degrees of freedom within the cluster including the spin-background variables. Especially for the irreducible one-quasiparticle hopping elements this leads to the observation, that clusters which do not host particles contribute due to a non-trivial spin-background on the cluster. Due to this spin-background the respective matrix element of the effective Hamiltonian might evaluate differently to the ground-state energy (which is subtracted to obtain the irreducible element).
\begin{figure}[ht]
	\centering
	\includegraphics[]{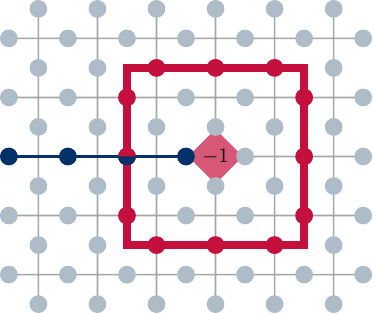}
	\caption{
		Example for a perturbation theory process which contributes to the excitation gap, although it does not directly act on the quasiparticle. The process consists of twelve $\sigma^x$ operators which act on the red sites forming a loop around the excitation and interacting with the string operator. This operator consists of $\sigma^z$ operators acting on the blue sites. Both operators, the loop and the string operator, act on the site which is red and blue.
		\label{fig::Loop}		  
	}
\end{figure}
A physically simple, but enlightening example is a cluster which winds around a charge as illustrated in Fig.~\ref{fig::Loop}. It is clear, that a process winding a flux around the charge results in a different contribution, than the same process without the charge even though the process does not directly involve the respective charge site.
So it is consistent that these clusters contribute to the irreducible hopping elements. Note that this type of process already occurs in order four perturbation theory.

\subsection{Identification of equivalent clusters}
\label{subsect::technicalities}
In the previous subsection we have shown that we can indeed write an expansion in the form of Eq.~\eqref{eq::linked_cluster1}. We also explained that for the ground-state energy and the one-quasiparticle matrix elements we can always choose a reference state, such that we can evaluate the contributions of the individual clusters without taking into account degrees of freedom which are not on the cluster. Throughout this subsection we assume to work in such a convention.

The next step is to exploit symmetries and structural equivalence to identify clusters which have the same contribution to the desired matrix element to arrive at an expansion of the form of Eq. \eqref{eq::linked_cluster2}. 
The distinction of clusters is based on the hypergraph decomposition explained in Ref. \cite{Muehlhauser2022}.
In contrast to the hypergraph decomposition in Ref.~\cite{Muehlhauser2022}, we have several types of sites in the present problem: spin-sites $s_i$, flux-sites $\bigtcplaq$, and charge-sites $\bigtcstar$.
From a technical point of view it is interesting that a given perturbation operator can act differently on different types of degrees of freedom, namely the spin-background variables and the eigenvalues of the star and plaquette operators. 
So while a cluster still consists of sites and bonds, the bonds have some orientation. This means different sites have different roles in the bonds.
While the eigenvalues of the stabilizer operators are always flipped, the action on the spin-background can be different.
\begin{figure}[ht]
	\begin{center}
		\includegraphics[]{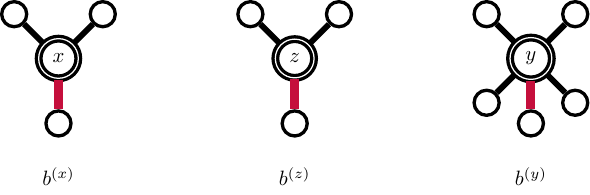}
	\end{center}
	\caption{The König representation of the different bonds with the contained sites is shown. The double circles represent the bonds, and the single circles represent the sites, namely the star and the plaquette eigenvalues and the spin background variable. Note the site corresponding to the spin-background variable is connected to the double circle by a thick red edge. \label{fig::single_bonds1}}
\end{figure}
So the structure of the bonds cannot be captured by ordinary hyperedges, which are just subsets of the vertex set, in a hypergraph representation of the cluster.  
Interestingly, several definitions of oriented hypergraphs can be found in the literature \cite{Mueller1981Oriented, Frankl1986OrientedUniform,Gallo1993Dirhyper, Rusnak2013, Bohmova2016, Ausiello2017Dirhyper, Jost2021}. Here, we assign labels to incident edge-vertex pairs, to include the roles of the vertices within the bonds.
Hypergraphs can be represented by bipartite graphs, which are also referred to as the König representation of a hypergraph \cite{Zykov1974, Konstantinova1995Molecular}.
The two parts of the bipartite König graph represent the edges (edge-part) and the vertices (vertex-part) of the hypergraph. Two vertices in the König representation are adjacent if and only if they correspond to an incident edge-vertex pair within the hypergraph.
Importantly, two hypergraphs are isomorphic if and only if their König representations are isomorphic \cite{Zykov1974, Konstantinova1995Molecular}.

The labels for the incident hyperedge-vertex pairs can be incorporated using edge colors within the König representation. Note that this is consistent with the fact that the edges in the König representation correspond to incidences of edge-vertex pairs in the original hypergraph.
Furthermore, we use vertex colors to distinguish the two parts of the bipartite graph and to incorporate the different bond-types into the edge-part of the König representation.

As an example we give the König representation of the clusters which consist of a single bond in Fig. \ref{fig::single_bonds1}.   
In the end this enhanced König representation is the basis for the distinction of equivalence classes of clusters within the full hypergraph decomposition presented here.
For calculations of the ground-state energy this representation of the clusters is indeed sufficient. For the hopping elements we need to explicitly incorporate the information on the involved states on the clusters, as the contribution of a clusters also depends on the states considered within the matrix element. Finally also this can be incorporated via vertex colors.

\subsubsection{Ground-state energy}

For the ground-state energy equivalent clusters are identified using hypergraph isomorphism accounting for the different roles of the vertices within the hyperedges and the different bond types $b^{(\alpha)}$ with $\alpha\in\{x,y,z\}$.
In terms of the König representation this means that the isomorphism preserves the edge and the vertex colors.
Note that one can reduce the König representation by omitting the vertices which represent the sites which are only contained in one bond within the cluster \cite{Muehlhauser2022}. The number of the omitted sites are easily deduced from the bond-types. 
We do not distinguish flux and charge-sites within the hypergraph representation,
because they are treated on equal footing in the operators $\tau_{n,b^{(\alpha )}}$ and also in the unperturbed Hamiltonian. So basically, the problem is decomposed into stabilizer eigenvalues which are relevant for the energetics, and spin-background variables, which we introduce to obtain the correct phases.
It is clear that the additional phase depends only on the action of the Pauli matrices on the reference state. This is correctly taken into account, once we know which Pauli matrices act on which spin-background variable and in which order.
From the edge-colors in the König representation it is clear, which sites correspond to the spin-background. The order is given by the sequence of operators $\tau_{n,b^{(\alpha )}}$.
Regarding the quasiparticles, the graph representation might result in graphs for which it is not clear whether vertices correspond to charge or plaquette sites. However, energetically this does not matter. 
In summary, we only separate the energetic degrees of freedom from the spin-background within the graph representation exploiting that elementary fluxes and charges have the same energy.

Another important question is how obtain the embedding factor $N(\mathcal{E},C)$. Typically, the ground-state energy is calculated per site, so $N(\mathcal{E},C)$ is the number of sub-clusters in $\mathcal{E}$ per spin-site. Here we calculate it per spin-site, i.e., modulo translational symmetry. 
To this end we count all connected sub-clusters of the system which contain a given spin-site $s_i$ and belong to $\mathcal{E}$.
Clearly, within this set there are still sub-clusters related by non-trivial translations. However, it is also clear that the number of translations of a sub-cluster which contain the spin-site $s_i$ id equal to the number of spin-sites within the sub-cluster. So in order to obtain $N(\mathcal{E},C)$, one can just count the number of sub-clusters in $\mathcal{E}$ which contain $s_i$, and divide by the number of sites of any sub-cluster in $\mathcal{E}$.
One can also require that for any two sub-clusters in $\mathcal{E}$ there exists an isomorphism taking the vertices representing the site $s_i$ to each other. Note that we still require $s_i$ to be on the sub-clusters. To obtain $N(\mathcal{E},C)$ the number of these sub-clusters is divided by the size of the orbit of the spin-site $s_i$ under the action of the automorphism group of the representing graph.

\begin{figure}
	\begin{center}
\includegraphics[]{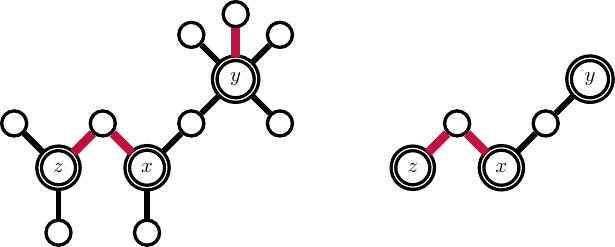}
	\end{center}
\caption{ {\textit Left}: Example for the König representation of a cluster.
The bonds are represented by double circles, where the bond types are explicitly given.
The different labels $x,y,z$ as well as double and single circles correspond to different vertex colors.
The vertices corresponding to spin-sites are explicitly distinguished as they are connected by red edges to the vertices representing the bonds. 
{\textit Right}: In the reduced König reprsentation the vertices from the vertex-part are omitted if they have degree one, reducing the complexity of the graphs representation.
}
\end{figure}

\subsubsection{Single-particle excitation gap}

\begin{figure}
	\begin{center}
	\includegraphics[]{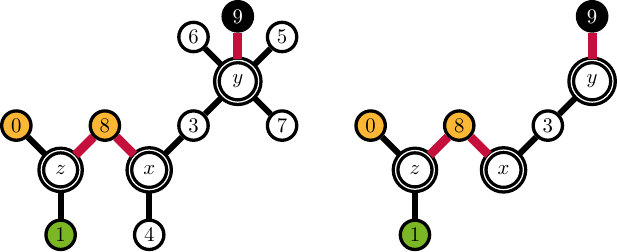}
	\end{center}
	\caption{\label{fig::HoppingKoenig}
	{\textit Left}: Illustration of the König representation for the hopping element
	$\bra{1 0 0 0 0 0 0 0; 1 1}
	H_{\rm eff}^c - E_0(c) 
	\ket{0 1 0 0 0 0 0 0;0 1}$.
	The $i$-th number in the states refers to the vertex with number $i$ in the König representation. Note that the enumeration of site vertices is not part of the actual representation, whereas the label of the bond-vertices is.
	We do not explicitly distinguish charge and spin sites, e.g., we cannot deduce which types of sites the vertices $5,6,7$ represent. The label $x,y,z$ of the bond vertices is part of the representation. {\textit Right}: By omitting the white site-vertices of degree one we again obtain a reduced representation.}

\end{figure}

For the single particle irreducible matrix elements only minor modifications are necessary in the graph representation. Recall that we are aiming to calculate irreducible matrix elements of the form 
	\begin{equation} a_{\vec{\delta}}^c = \bra{\vec{r}_2} H_{\rm{eff}}^c - E_0(c) \ket{\vec{r}_1}
\end{equation}
with $\vec{\delta}=\vec{r}_2-\vec{r}_1$. In Subsect.~\ref{subsect::individual_clusters} we established that using appropriate conventions for the calculations we can execute the calculations on finite clusters, and they are not affected by any degrees of freedom outside the cluster. 
So we can encode the product state on the cluster using vertex colors for all vertices which represent sites. Note that these vertices must still be distinguishable from the vertices which represent bonds.
Depending on the local state at a given site in the states $\ket{\vec{r}_1}$ and $\ket{\vec{r}_2}$ involved in the concrete matrix element we assign colors in the following way
\begin{equation}
	\label{eq:local_states}
\begin{blockarray}{ccc}
&\ket{0}& \ket{1}  \\
\begin{block}{c[cc]}
\bra{0}& \begin{tikzpicture}
\node[draw, circle, inner sep = 1.5mm, ultra thick]{};
\end{tikzpicture}&\begin{tikzpicture}
\node[draw, circle, inner sep = 1.5mm, ultra thick, fill = FAUgelb]{};
\end{tikzpicture}\\
\bra{1}&
\begin{tikzpicture}
\node[draw, circle, inner sep = 1.5mm, ultra thick, fill = FAUgruen]{};\end{tikzpicture}& \begin{tikzpicture}\node[draw, circle, inner sep = 1.5mm, ultra thick, fill = black]{};\end{tikzpicture}\\
\end{block}
\end{blockarray}
\end{equation}
where the states label the values of $\tilde{x}_\tcstar$, $\tilde{z}_\tcplaq$ or $\tilde{s}_i$ at the respective site. A white filling of the vertex means, that the eigenvalue associated with the site is zero within both states $\ket{\vec{r}_1}$ and $\ket{\vec{r}_2}$, while an orange vertex indicates that the respective eigenvalue changes from $1$ to $0$. A specific example for this representation is shown in Fig.~\ref{fig::HoppingKoenig}. In summary we distinguish seven different types of vertices with different colors: Three different types of vertices representing bonds $b^{(x)},b^{(y)},b^{(z)}$ and four different types of site vertices representing changes of the local states as indicated in Eq.~\eqref{eq:local_states}.

Next, we adress the question, which sub-clusters we need to consider. The obvious answer is every sub-cluster $c$ with 
\begin{equation} \bra{\vec{r}_2}  H_{\rm eff}^c  \ket{\vec{r}_1} \neq \bra{\vec{r}_2}  E_0(c)  \ket{\vec{r}_1}. \end{equation}

\begin{figure}[htb]
	\begin{center}
	\includegraphics[]{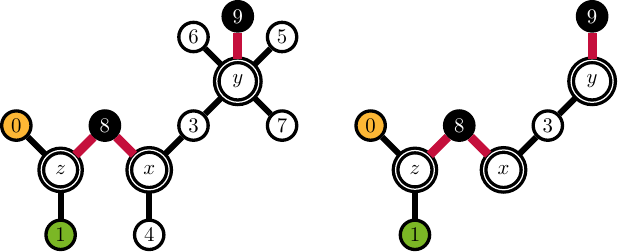}
		
	\end{center}
		\caption{\label{fig::HoppingKoenig2}
			Illustration of the König representation of a hopping element, without taking into account the spin-background of the state $\ket{\vec{r}_2}$.
			In contrast to Fig. \ref{fig::HoppingKoenig} the vertex $8$ is now black instead of orange. 
}
\end{figure}

\begin{figure}[htb]
	\begin{center}
		\includegraphics[]{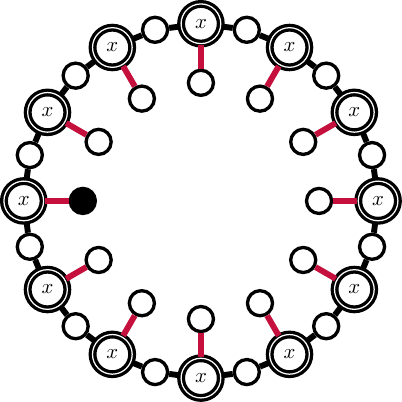}
	\end{center}
	\caption{A graph which can only contribute to diagonal matrix elements $\bra{\vec{r}, \bigtcstar}  H_{\text{eff}} - E_0\ket{\vec{r}, \bigtcstar}$.
	The spin-background distinguishes this graph, representing a cluster which encloses a charge from a graph which represents a cluster which does not enclose a charge.
	\label{fig::HoppingKoenig3} 
	}
\end{figure}

First of all, it is clear that a sub-cluster can only contribute if it contains all stabilizer sites, which change their local state within the matrix element.
Next, the sub-cluster should host excitations or the initial state $\ket{\vec{r}_1}$
should have a non-trivial spin-background. 
Interestingly the spin-background of $\ket{\vec{r}_2}$ does not matter. 
In order to determine the state $\ket{\rm f}$ resulting from the action of a sequence of operators $\tau_{n,b^{(\alpha)}}$ on a state $\ket{\vec{r}_1}$ the spin-background of the state $\ket{\vec{r}_1}$ is crucial.
Using an appropriate reference state the product of operators $\tau_{n,b^{(\alpha)}}$ yields the result in terms of a phase factor $[\mathrm{i}^{n_a}]$, and a state equivalent to a canonical state $\ket{\vec{r}^{\prime \prime}}$. So the spin-background of the state $\ket{\vec{r}_2}$ is not important, as 
	\begin{equation}
		\braket{\vec{r}_2 \vert \rm f} =  \bra{\vec{r}_2} [\mathrm{i}^{n_a}] \ket{ \vec{r}^{\prime\prime}} = [\mathrm{i}^{n_a}] \cdot \delta_{\vec{r}_2,\vec{r}^{\prime\prime}}\,.
	\end{equation}
This also implies that the spin-background of $\ket{\vec{r}_2}$ does not play a role in the representation and the selection of the sub-clusters. 
As a consequence, we can ignore the spin-background of $\ket{\vec{r}_2}$ for the graph representation, while we still need to include the spin-background of $\ket{\vec{r}_1}$ within the graphs. Accordingly, for vertices representing spin-background variables we only use two colors, based on the respective eigenvalue in the state $\ket{\vec{r}_1}$
\begin{equation}
\begin{blockarray}{ccc}
&\ket{0}& \ket{1}  \\
\begin{block}{c[cc]}
\bra{0}& \begin{tikzpicture}
\node[draw, circle, inner sep = 1.5mm, ultra thick]{};
\end{tikzpicture}&\begin{tikzpicture}
\node[draw, circle, inner sep = 1.5mm, ultra thick, fill = black]{};
\end{tikzpicture}\\
\bra{1}&
\begin{tikzpicture}
\node[draw, circle, inner sep = 1.5mm, ultra thick]{};\end{tikzpicture}& \begin{tikzpicture}\node[draw, circle, inner sep = 1.5mm, ultra thick, fill = black]{};\end{tikzpicture}\\
\end{block} 
\end{blockarray}.
\end{equation}
With this the König representation from Fig.~\ref{fig::HoppingKoenig} is replaced by the representation in Fig.~\ref{fig::HoppingKoenig2}. For the stabilizer degrees of freedom we keep the convention defined in Eq. \eqref{eq:local_states} so we still distinguish seven different types of vertices. This way of coloring the vertices has the advantage that the rules which clusters need to be taken into account are formulated more easily. For off-diagonal elements $\bra{\vec{r}_2, \bigtcstar}  H_{\text{eff}} - E_0\ket{\vec{r}_1, \bigtcstar}$ with $\vec{r}_2 \neq \vec{r}_1$ all sites represented by green or orange vertices need to be included in a cluster, otherwise its reduced contribution vanishes. Instead, for the diagonal element the reduced contribution of a cluster is zero, if no site represented by a black vertex is contained in the cluster.
A specific cluster on which the winding process of a flux around a charge in Fig.~\ref{fig::Loop} occurs is represented by the graph in Fig.~\ref{fig::HoppingKoenig3}. While this graph does not contain any vertices corresponding to excitations the non-trivial spin-background suffices to take the effects of the charge (which is not on the cluster) into account. Note that this graph can only play a role within matrix elements $\bra{\vec{r}_2, \bigtcstar}  H_{\text{eff}} - E_0\ket{\vec{r}_1, \bigtcstar} $ if $\vec{r}_2 = \vec{r}_1$.
Note that also clusters which do not enclose a charge can have a non-trivial spin-background in such a hopping element. Currently, we simply consider the respective equivalence classes within the calculations. It is interesting whether a scheme to sort out these clusters can be devised and how it affects the efficiency of the method.  

\subsection{Series for ground-state energy and excitation gaps}
Using the described hypergraph decomposition, we calculated the ground-state energy per spin up to order 10 using the pCUT method. 
The ground-state energy, which is symmetric under the exchange of $h_x$ and $h_z$ due to self duality, can be written down by summarizing certain terms using the notation $S_j=h_x^j+h_z^j$ and $P_{2j}=h_x^jh_z^j$, which has been introduced in \cite{KamforDiss}. It reads 
\begin{align}
e_0=&-1/2-\frac{1}{2}S_2 -\frac{1}{4}h_y^2-\frac{15}{8}S_4-\frac{7}{32}S_2h_y^2+\frac{1}{4}P_4-\frac{13}{192}h_y^4-\frac{147}{8}S_6-\frac{371}{128}S_4h_y^2\nonumber\\
&+\frac{113}{32}P_4S_2-\frac{1045}{3456}S_2h_y^4+\frac{2003}{384}P_4h_y^2-\frac{197}{3072}h_y^6-\frac{18003}{64}S_8-\frac{1954879}{36864}S_6h_y^2\nonumber\\
&+\frac{6685}{128}P_4S_4-\frac{34054175}{3981312}S_4h_y^4+\frac{146861}{2304}P_4S_2h_y^2-\frac{15343549}{26542080}S_2h_y^6+\frac{20869}{384}P_8\nonumber\\
&+\frac{5020085}{497664}P_4h_y^4-\frac{163885}{1769472}h_y^8-\frac{5420775}{1024}S_{10}-\frac{1563459523}{1327104}S_8h_y^2+\frac{39524033}{36864}P_4S_6\nonumber\\
&-\frac{1115105409427}{5733089280}S_6h_y^4+\frac{10058235445}{7962624}P_4S_4h_y^2-\frac{4219640835497}{191102976000}S_4h_y^6\nonumber\\
&+\frac{5650925}{6912}P_8S_2+\frac{20854097563}{143327232}P_4S_2h_y^4-\frac{483890940281}{382205952000}S_2h_y^8+\frac{1202498305}{1990656}P_8h_y^2\nonumber\\
&+\frac{1994817656221}{71663616000}P_4h_y^6-\frac{186734746441}{1146617856000}h_y^{10}.
\label{eq:gs_square}
\end{align}
The one-particle excitation energies have been calculated in order 9 for arbitrary field directions and in order 10 for a parallel field direction with $h_y=0$. Here we only focus on the charge and flux gap which is always located at $\vec{k}=(0,0)$.
Due to self-duality, we can constrain ourselves to the charge gap $\Delta^{\tcstar}$ which reads for arbitrary field direction
\begin{align}
\Delta^{\tcstar}=&1-4 h_z -4 h_z^{2}-h_y^{2}-12 h_z^{3}+\frac{11}{4} h_y^{2} h_z +2 h_x^{2} h_z -36 h_z^{4}-9 h_y^{2} h_z^{2}-\frac{15}{16} h_y^{4}+3 h_x^{2} h_z^{2}
\nonumber\\
&+17 h_x^{2} h_y^{2}+5 h_x^{4}-176 h_z^{5}+\frac{473}{64} h_y^{2} h_z^{3}+\frac{17}{4} h_y^{4} h_z +\frac{83}{4} h_x^{2} h_z^{3}+\frac{9}{4} h_x^{2} h_y^{2} h_z +\frac{27}{2} h_x^{4} h_z 
\nonumber\\
&-\frac{2625}{4} h_z^{6}-\frac{7971}{64} h_y^{2} h_z^{4}-\frac{135619}{3456} h_y^{4} h_z^{2}-\frac{575}{384} h_y^{6}+63 h_x^{2} h_z^{4}+\frac{1305}{8} h_x^{2} h_y^{2} h_z^{2}
\nonumber\\
&+\frac{13621}{1152} h_x^{2} h_y^{4}+71 h_x^{4} h_z^{2}+\frac{14267}{96} h_x^{4} h_y^{2}+92 h_x^{6}-\frac{14771}{4} h_z^{7}-\frac{238621}{1152} h_y^{2} h_z^{5}
\nonumber\\
&+\frac{1782929}{20736} h_y^{4} h_z^{3}+\frac{799973}{110592} h_y^{6} h_z +\frac{28633}{64} h_x^{2} h_z^{5}+\frac{13807}{48} h_x^{2} h_y^{2} h_z^{3}-\frac{3031}{13824} h_x^{2} h_y^{4} h_z 
\nonumber\\
&+\frac{925}{4} h_x^{4} h_z^{3}+\frac{1142149}{4608} h_x^{4} h_y^{2} h_z +\frac{495}{2} h_x^{6} h_z -\frac{940739}{64} h_z^{8}-\frac{4663837}{1728} h_y^{2} h_z^{6}
\nonumber\\
&-\frac{1760584999}{1990656} h_y^{4} h_z^{4}-\frac{1495320677}{19906560} h_y^{6} h_z^{2}-\frac{26492351}{7962624} h_y^{8}+\frac{118029}{64} h_x^{2} h_z^{6}
\nonumber\\
&+\frac{5186533}{1728} h_x^{2} h_y^{2} h_z^{4}+\frac{24547709}{165888} h_x^{2} h_y^{4} h_z^{2}+\frac{98263727}{3981312} h_x^{2} h_y^{6}+\frac{19263}{16} h_x^{4} h_z^{4}
\nonumber\\
&+\frac{2199571}{4608} h_x^{4} h_y^{2} h_z^{2}+\frac{3032191}{31104} h_x^{4} h_y^{4}+\frac{80999}{96} h_x^{6} h_z^{2}+\frac{7715431}{3072} h_x^{6} h_y^{2}+\frac{35649}{16} h_x^{8}
\nonumber\\
&-\frac{11472297}{128} h_z^{9}-\frac{4691521349}{442368} h_y^{2} h_z^{7}+\frac{538730849}{497664} h_y^{4} h_z^{5}+\frac{261699729407}{2388787200} h_y^{6} h_z^{3}
\nonumber\\
&+\frac{43413180467}{2388787200} h_y^{8} h_z +\frac{14650547}{1152} h_x^{2} h_z^{7}+\frac{2099164741}{165888} h_x^{2} h_y^{2} h_z^{5}-\frac{215708093}{746496} h_x^{2} h_y^{4} h_z^{3}
\nonumber\\
&+\frac{41328278581}{597196800} h_x^{2} h_y^{6} h_z +\frac{918461}{144} h_x^{4} h_z^{5}+\frac{1791338425}{331776} h_x^{4} h_y^{2} h_z^{3}-\frac{1255335587}{11943936} h_x^{4} h_y^{4} h_z 
\nonumber\\
&+\frac{18372481}{4608} h_x^{6} h_z^{3}+\frac{2913725647}{663552} h_x^{6} h_y^{2} h_z +\frac{162525}{32} h_x^{8} h_z\,.
\label{eq:gap}
\end{align}
Due to reduced computational time, in the parallel field case with $h_y=0$, the charge gap was calculated up to order 10
\begin{align}
\Delta^{\tcstar}=&1-4 h_z -4 h_z^{2}-12 h_z^{3}+2 h_x^{2} h_z -36 h_z^{4}+3 h_x^{2} h_z^{2}+5 h_x^{4}-176 h_z^{5}+\frac{83}{4} h_x^{2} h_z^{3}\nonumber\\
&+\frac{27}{2} h_x^{4} h_z -\frac{2625}{4} h_z^{6}+63 h_x^{2} h_z^{4}+71 h_x^{4} h_z^{2}+92 h_x^{6}-\frac{14771}{4} h_z^{7}+\frac{28633}{64} h_x^{2} h_z^{5}+\frac{925}{4} h_x^{4} h_z^{3}\nonumber\\
&+\frac{495}{2} h_x^{6} h_z -\frac{940739}{64} h_z^{8}+\frac{118029}{64} h_x^{2} h_z^{6}+\frac{19263}{16} h_x^{4} h_z^{4}+\frac{80999}{96} h_x^{6} h_z^{2}+\frac{35649}{16} h_x^{8}\nonumber\\
&-\frac{11472297}{128} h_z^{9}+\frac{14650547}{1152} h_x^{2} h_z^{7}+\frac{918461}{144} h_x^{4} h_z^{5}+\frac{18372481}{4608} h_x^{6} h_z^{3}+\frac{162525}{32} h_x^{8} h_z\nonumber\\
&-\frac{287258435}{768} h_z^{10}+\frac{96935975}{1728} h_x^{2} h_z^{8}+\frac{107740069}{3456} h_x^{4} h_z^{6}+\frac{21893537}{1536} h_x^{6} h_z^{4}\nonumber\\
&+\frac{43593271}{3072} h_x^{8} h_z^{2}+\frac{1873147}{32} h_x^{10}\,.
\end{align}
The flux gap $\Delta^{\tcplaq}$ can be again obtained by exchanging $h_x$ and $h_z$ in these expressions. These results agree with previous high order series expansions \cite{Dusuel2011,KamforDiss}. In both the arbitrary and parallel field cases we were able to exceed the previously known results by one order.

\section{Conclusions}
\label{sect::conclusions}
In this work we presented how to treat non-local anyonic statistics with a high-order linked-cluster expansions using a full hypergraph decomposition. This  we exemplified for the topological phase of the toric code in the presence of a uniform field possessing Abelian charge and flux quasi-particles. Technically, we used the pCUT method to obtain series expansions for the ground-state energy and the one-quasiparticle charge and flux gap. Note that also other perturbative techniques like Takahashi's perturbation theory \cite{Takahashi1977, Klagges2012} or Löwdin's partitioning technique \cite{Loewdin1962, Kalis2012} can be applied to calculate these properties of the perturbed topological phase. 

The improvement in terms of the achieved maximal perturbative order compare to calculations done with Entings finite lattice method \cite{Vidal2009, Vidal2011, Dusuel2011} is only moderate. Nevertheless, the explicit treatment of the non-locality of the fractional statistics of charges and fluxes within a full hypergraph decomposition is an asset on its own. Further, there will be a benefit of a full hypergraph decomposition when performing similar calculations for three-dimensional perturbed topological phases like the ones in the three-dimensional toric code \cite{Hamma2005, Nussinov2008,  Reiss2019} or the fractonic X-Cube model \cite{Vijay2016} in the presence of external perturbations. Finally, while we expect that extensions to other Abelian topological phases can also be treated in the described framework, extensions to perturbed non-Abelian topological phases pose an interesting and open challenge.   

\section*{Acknowledgements}

\paragraph{Funding information}
KPS acknowledges financial support by the German Science Foundation (DFG) through the grant SCHM 2511/11-1 as well as gratefully acknowledge the support by the Deutsche Forschungsgemeinschaft (DFG, German Research Foundation) - Project-ID 429529648-TRR 306 QuCoLiMa ("Quantum Cooperativity of Light and Matter") as well as the Munich Quantum Valley, which is supported by the Bavarian state government with funds from the Hightech Agenda Bayern Plus.

\begin{appendix}
	\section{Identifying non-contributing clusters or isomorphism classes}
		\label{sect::non_contributing}
		The selection rules given in \cite{He1990, Oitmaa2006} can be adapted for the problem presented in this paper. We will shortly describe how this can be done but refer to \cite{Muehlhauser2022} for a longer discussion in the context of hypergraph expansions. 
		We will explain the heuristics for the vacuum. A generalization for hopping elements is straightforward.
		
		\subsection{Vertex Degree}
			We are considering the degrees of the vertices which represent stabilizer operators.
			Any such vertex which has an odd degree within the full König representation induces
			that the perturbation has to act more than once on an adjacent bond. Of course such a vertex can share this bond with other vertices of odd degree, though the maximum number of sites in a bond $c_{\text{max}}$ poses an upper limit to this. Here $c_{\text{max}} = 4$ for the general field and $c_{\text{max}} = 2$ for a parallel field ($h_y=0$).
			So having $m$ vertices of odd degree which represent a stabilizer eigenvalue we need to act
			at least $\text{ceil}(m/c_{\text{max}})$ additional times with the perturbation, where $\text{ceil}$ denotes rounding up to the next integer. For a cluster $C$ with $n_b
			$ bonds and $m$ such vertices any process contributing to $\kappa(C)$ has to act at least $ n_b + \text{ceil}(m/c_{\text{max}})$.
			So at perturbation order $k$ the contribution $\kappa(C)$ of a cluster $C$ will be zero if
		\begin{equation}
		c_{\text{max}} (k-n_b) < m. \label{eq::dtlim1} 
		\end{equation}	
		It is obvious, that if a cluster fulfills \eqref{eq::dtlim1} all clusters created adding further bonds will fulfill the inequality as well.
		
		This procedure is applicable to clusters and isomorphism classes as well. It is very powerful because it prunes whole subtrees during the search for new clusters.
		On the cluster level this scheme can be extended taking into account the actual bonds of the system. This can be achieved by storing whether two sites are contained in a at least one of the bonds simultaneously for any pair of sites. Note that here we are considering all the bonds in the entire system. Once we have the vertices of odd degree which represent the stabilizer eigenvalues, we know which of these vertices have to be in different bonds. As an example suppose we have four such vertices in a cluster $C$, and all of these vertices have to be in different bonds. With this we mean, that there is no bond containing two of these vertices in the entire system (including also the bonds in $C$). Then from the simple estimate above we conclude that the cluster contributes in order $n_b+1$ (considering $h_y \neq 0 $), where again $n_b$ is the number of bonds in the cluster $C$. Instead, with the refined estimate we conclude that it contributes only in order $n_b+4$. How one can estimate a lower bound for the number of different bonds needed using graph coloring algorithms is explained in \cite{Muehlhauser2022}.
		Let us conclude with the remark that also this estimate can be used to prune entire branches of the search tree. It is advisable to use both of these techniques during the generation of the clusters, in order to improve the performance of the cluster generation.
		
		\subsection{Doubling Edges}	
		Another possibility to sort out non-contributing isomorphism classes is a technique presented by He et al. \cite{He1990}. We consider a cluster $C$ representing a given isomorphism class and try to obtain a cluster where all vertex degrees (again here we consider only vertices which represent stabilizer degrees of freedom) are even by duplicating bonds within the cluster $C$.
		The smallest number of bonds of such a cluster is the order in which the isomorphism class can contribute.
		We suggest applying this technique after the cluster generation on the level of the isomorphism classes. This reduces the number of considered equivalence classes further and avoids superfluous calculations.
	
\end{appendix}

\nolinenumbers

\end{document}